# Optical Quasiparticles: Skyrmions, Bimerons and Skyrmionic Hopfions in Paraxial Laser Beams

Srinivasa Rao Allam[1,2,3*], Yuto Yoneda[1], and Takashige Omatsu[1,2]
[1]Graduate School of Engineering, Chiba University, Chiba, 263-8522, Japan
[2]Molecular Chirality Research Centre, Chiba University, Chiba, 263-8522, Japan
[3]Institute for Advanced Academic Research, Chiba University, Chiba, 263-8522, Japan
[*]asvrao@chiba-u.jp, sri.jsp7@gmail.com

**Abstract**

Skyrmions, merons, bimerons, and skyrmionic hopfions are quasiparticles which have stable topological textures. They have been observed across multiple physical domains including nucleons, water waves, magnetic materials, electromagnetic fields, condensed-matter physics, liquid crystals, and Bose-Einstein condensates. The nontrivial topological texture of these quasiparticles exhibited promising applications across several research fields including cosmology, particle physics, superfluids, high-energy physics, optics, condensed matter, early-universe cosmology, cold quantum matter, and liquid crystals. These quasiparticles have most recently been observed in paraxial vector beams. Due to the inhomogeneous polarization distribution of vector beams, the quasiparticle textures reside in the Stokes vector fields. Here, we review recent developments in paraxial quasiparticle research and provide detailed information on their fundamental properties and future prospects. We discuss various techniques which have been used to experimentally generate these quasiparticles and we examine a number of their proposed applications. Notably, this review offers insight into the concepts and techniques which can be applied to the generation of optical skyrmions in paraxial laser beams, and investigates their intriguing applications across both fundamental and applied optics.

## 1. Introduction

A quasiparticle is comprised of a set of particles taken from a single group within a physical system, and these particles' collective behavior is treated as a single particle. These quasiparticles can have localized topologically protected structures in two-dimensional (2D) and three-dimensional (3D) spaces which can be characterized by respective 2D and 3D topologies. Among these quasiparticles, skyrmions, merons, bimerons, and hopfions have garnered particular interest within scientific domains including cosmology, particle physics, superfluids, high-energy physics, condensed matter, early-universe cosmology, cold quantum matter, and liquid crystals. Skyrmions in 3D space were first proposed in 1961 as a topological model for nuclei by Tony Skyrme. As detailed in his seminal paper, a skyrmion is considered a field solution of the equation which minimizes the energy of the system. The concept was utilized to explain the interactions between mesons and baryons [1]; following this, further developments in topological models explored the use of other quasiparticles. The topological textures of these quasiparticles have been observed in the field vectors used across several physical domains, a number of which are listed in Table 1.

Table 1: Quasiparticles observed in different areas of research.

| Field vector | Physical domain |
|---|---|
| Magnetization field | Magnetism [2] |
| Photonic spin | Spin-orbit coupling of focused vector beam [3] |
| Electric field | Evanescent waves [4] |
| Stokes vector | Paraxial Poincaré beams [5] |
| Spin of thermally excited quasielectron-quasihole pairs | Quantized Hall effect [6] |
| The combination of phase field and vector potential in electrodynamics | Superconductors [7] |

| Liquid crystal orientation | Liquid crystals [8] |
|---|---|
| spin-1/2 condensate of $^{87}$Rb atoms | Bose–Einstein condensates [9] |
| Velocity fields | Sound waves [10] |
| Pseudospin vectors | Photonic crystals [11] |
| Surface plasmon polariton | Metal film [12] |
| Photonic spin angular momentum | Meta-surface [13] |
| Electromagnetic vectors | Super toroidal pulses [14] |
| Electric field | Free space $4\pi$ focusing system [15] |
| Water waves | Water surface [16] |
| Momentum vector | Mie Scattering Fields [17] |
| Pointing vector | Counter propagating vector beams [18] |

Among all proposed quasiparticles which include skyrmions, merons, bimerons, skyrmioniums, bimeroniums, and skyrmionic hopfions, the understanding of the topological texture of skyrmions is perhaps the most well-developed, and can be applied to the analysis of other quasiparticles. The topological texture of skyrmions is confined within a single plane (2D) and are referred to as 2D skyrmions (and sometimes 'baby skyrmions') to distinguish them from 3D topological textures. Here, the spin of a topological texture is a normalized vector in 3D which is distributed over a 2D plane. All the vectors located at the boundary are oriented in the same direction but opposite to the vector located at the center. The orientation of the vector from the center to the boundary is smoothly connected with a uniform transformation. The orientation of vectors in this uniform transformation are not unique, but rather, completely depend on the transverse phase profile of superposed spin vectors. Irrespective of their order, these topological textures, which are based on their field vector orientation, are broadly classified into Néel-type, Bloch-type, and anti-type skyrmions. Néel-type and Bloch-type skyrmions exhibit hedgehog and vortex textures respectively, while the anti-skyrmions exhibit a saddle texture. As shown in Fig. 1, the skyrmionic textures can be stereographically projected from a 2D plane to a 2-sphere. The vector field at the center of the skyrmion is mapped to the south pole while the peripheral vectors are projected onto a single point on the north pole. The vectors present at the south and north poles are oriented in opposite directions. In this sense, as one moves radially along the 2D plane, each vector is projected on to the sphere with increasing latitude without changing the orientation of the vector.

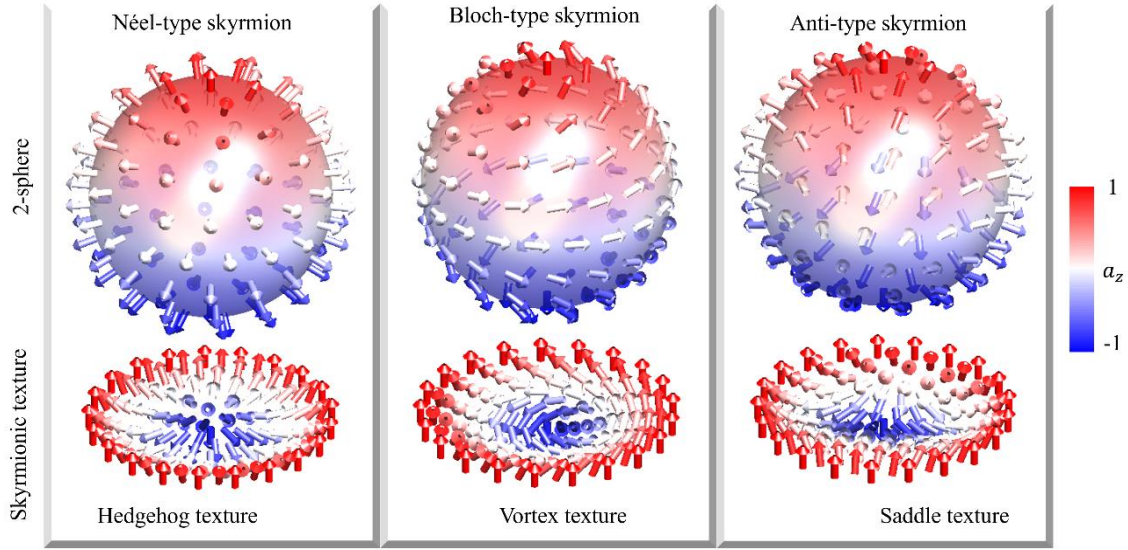

Fig.1. Images showing the stereographic projection of the topological textures of first-order skyrmions from a 2D plane onto a 2-sphere.

The aim and motivation of this topical review are to explore the fundamental properties of optical quasiparticles in paraxial laser beams and examine how they may be applied to applications in modern science and technology. Specifically, the quasiparticles discussed in this review are skyrmions, bimerons, skyrmioniums, bimeroniums, and skyrmionic hopfions. This review is divided into multiple sections. Following the introduction to quasiparticles in section 1, in section 2 we provide a comparative study of magnetic skyrmions generated in electronic paraxial beams and optical skyrmions generated in paraxial laser beams. Particular focus is given to the discussion of the polarization distribution and topological textures of skyrmions. Skyrmionic textures are investigated within the context of Laguerre Gaussian (LG) and Bessel beams via a detailed literature survey. In section 3, we provide a detailed discussion of how meron lattices and bimerons are produced in vector laser beams. In addition to skyrmion, and bimeron textures, we also briefly discuss skyrmioniums, and bimeroniums in section 4. In section 5, the topological texture and hopf fibration of skyrmionic hopfions are discussed in terms of polarization. We also detail the necessary conditions required for the experimental generation of skyrmionic hopfions. The experimental realization of 2D optical quasiparticles in paraxial laser beams (which are discussed in sections 2-5), is investigated in section 6. In section 7, we briefly discuss how to tune the wavelength and topological texture of optical quasiparticles through nonlinear wave-mixing. We also discuss possible experimental artifacts in the design and characterization of optical quasiparticles in section 8. In section 9, we discuss existing and emergent applications of paraxial laser beams; and a summary and conclusion are given in section 10.

## 2. Skyrmions

Recently, magnetic skyrmions have garnered significant attention owing to their ultra-compact size, topologically protected stability, and low current requirements [2,19-22]. Photonic counterparts of magnetic skyrmions have been observed and studied across multiple disciplines of optics [4, 23, 24]. In recent years, these textures have been observed as Stokes vectors in Poincaré beams [5,25,26] and have been referred to as optical skyrmions in paraxial laser beams (paraxial skyrmions) or skyrmionic beams. As depicted in Fig. 2, magnetic skyrmions and optical skyrmions, under the paraxial approximation, share a number of similarities (more details can be found in [5]). In these fields, the skyrmionic state can be obtained by the superposition of orthogonal wave functions derived from a paraxial beam of either light [27] or electrons [28]. The superposition state of a skyrmion takes the form

$$|F\rangle = \frac{1}{\sqrt{2}}[f_0|0\rangle + exp(i\theta_0)f_1|1\rangle]. \qquad (1)$$

Where $f_0$ and $f_1$ are two orthogonal spatial modes and the states $|0\rangle$ and $|1\rangle$ are orthogonal basis vectors in 2D space. The basis vectors correspond to the polarization basis of the laser beam and the spin basis in the electron beam. The global phase difference between the two modes is $\theta_0$. The skyrmion field vectors of these beams are obtained by the inner product of the state given by Eq. 1 on the vector operator created by Pauli matrices. The superposition state can thus produce all types of skyrmionic textures with the order tunable by changing the amplitude of the spatial modes and interchanging the state vectors. The skyrmionic textures created in the electron beam can encompass all possible spin states of the Bloch sphere. Hence, the magnetic skyrmion texture can be projected onto the Bloch sphere [Fig. 2(*a*)]; and thus the skyrmionic field vector can be considered equivalent to the Bloch vector. When it comes to light, we can produce all the polarization states of the Poincaré sphere [29] within the beam cross-section of paraxial laser beams which we call Poincaré beams [30,31]. The inhomogeneous polarization distribution of Poincaré beams has a skyrmionic texture described in terms of Stokes vectors. Thereby, we can project the skyrmionic texture of paraxial laser beams onto the Poincaré sphere [Fig. 2(*b*)]. In this case, the skyrmionic field vector can be considered equivalent to the Stokes vector.

Enthusiasm and interest in skyrmionic textures created in paraxial laser beams has accelerated due to their unique characteristics. Quasiparticles observed across most fields of science are limited to specific types of skyrmionic textures. However, in the case of paraxial laser beams, there is the capacity to easily realize all types of quasiparticles with tunable order. Paraxial light skyrmions can also propagate in free space and as such, offer an additional degree of freedom which may be exploited in applications. Also, their stable topological texture (created by their non-uniform polarization distribution) can further enhance applications which leverage polarization-sensitive light-matter interactions.

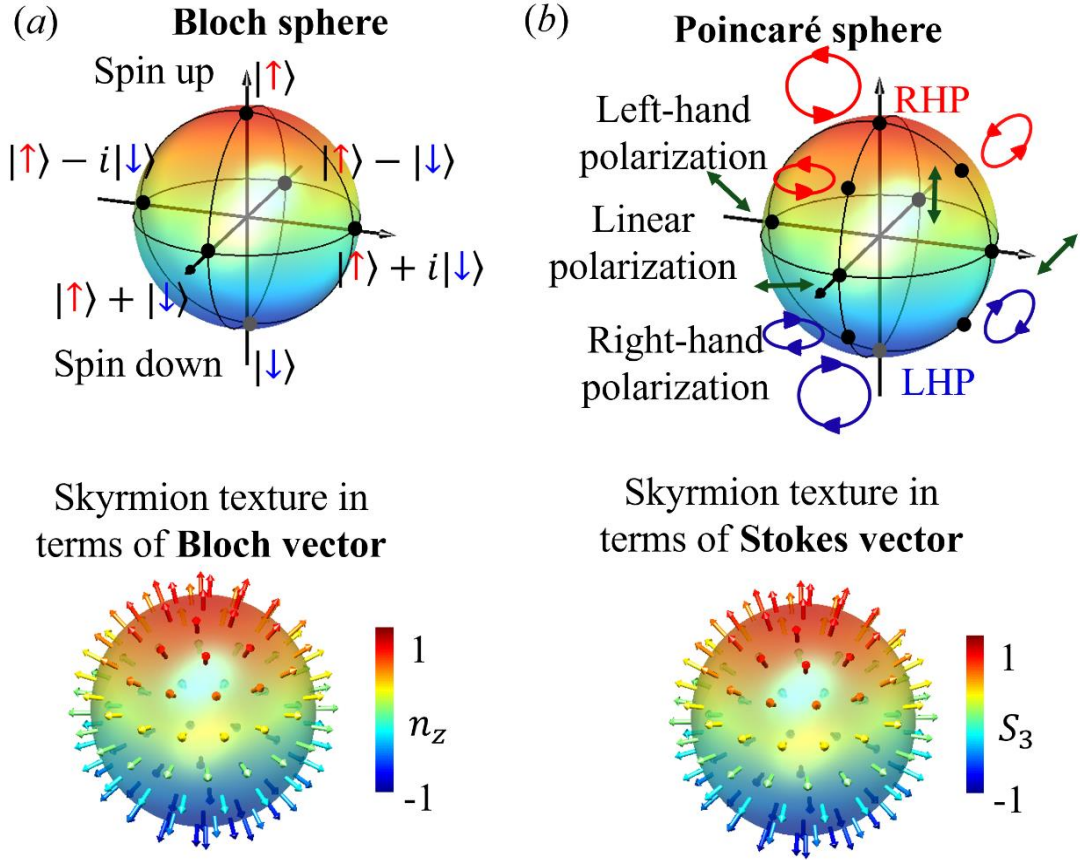

Fig. 2. Images showing the comparison between skyrmions generated in electron spin and light polarization. RHP is right-handed polarization and LHP is left-handed polarization.

The controlled non-uniform polarization distribution created in vector laser beams can allow us to sculpt all types of textures of skyrmions with order tunability. The topological characteristics of skyrmionic textures results from nonuniform polarization across their beam cross-section, which in and of itself, is attributed to the radially-dependent intensity distribution and azimuthally-dependent phase of superposed laser beams. As such, skyrmions can be produced through the use of Laguerre-Gaussian (LG) and Bessel beams. These two types of laser beams have both an order-dependent radial intensity distribution and an azimuthal phase gradient with a cylindrical symmetry. These beams are also Eigenmodes of a laser cavity and as such, preserve their shape when propagating in free-space. Hence, topological textures created using these paraxial laser beams also preserve their topology while propagating.

## 2.1. Skyrmions in paraxial LG beams

The propagation-dependent transverse profile of a paraxial LG mode ($LG_p^\ell$) with a radial index $p$ and an azimuthal index (also referred to as topological charge, winding number, or orbital angular momentum (OAM) number) $\ell$ in the mathematical form is given by [32,33]

$$LG_p^\ell(r,\phi,z) = \sqrt{\frac{p!}{\pi(|l|+p)!}} \frac{\sqrt{2}^{|\ell|+1} r^{|\ell|}}{w^{|\ell|+1}} L_p^{|\ell|}\left(\frac{2r^2}{w^2(z)}\right) e^{-\frac{r^2}{w^2(z)}} e^{i\ell\phi + i\xi - i\frac{kr^2}{2R(z)}}, \qquad (2)$$

where $r$ and $\phi$ are the radial and azimuthal coordinates, $k$ is the wave number, $L_p^{|\ell|}$ is the associated Laguerre polynomial, $w(z) = w_0\sqrt{1+z^2/z_R^2}$ is the beam spot size at arbitrary longitudinal position $z$, $w_0$ is the Gaussian beam waist at $z = 0$, and $z_R$ is the Rayleigh length, respectively. The radius of curvature of the mode is given by $R(z) = z\sqrt{1+z_R^2/z^2}$. The Gouy phase $\xi = (1 + |\ell| + 2p)\tan^{-1}\left(\frac{z}{z_R}\right)$ is a propagation-dependent phase and depends on the radial and azimuthal indices of the LG beam. The Poincaré beam can be obtained by the superposition of two collinear, co-propagating orthogonal LG beams present within the circular polarization (CP) basis. The state of the Poincaré beam in terms of LG modes is given by

$$|\Psi_P(r,\phi,z)\rangle = \frac{1}{\sqrt{2}}\left[LG_{p_R}^{\ell_R}(r,\phi,z)|R\rangle + exp(i\theta_0)LG_{p_L}^{\ell_L}(r,\phi,z)|L\rangle\right], \tag{3}$$

here $|R\rangle$ and $|L\rangle$ represent the right and left CP states. The order-dependent transverse intensity and phase of LG modes are illustrated in Fig. 3. The radius of the annular intensity distribution and the azimuthal gradient phase of an LG mode increases with its order [34]. When we superpose two orthogonal LG modes present in orthogonal CP states, the azimuthal phase difference gradient created between the LG modes provides a uniform change in the azimuth of the polarization, and the radial dependent intensity distribution of individual LG modes produces a continuous change from one hand polarization to other hand polarization in the radial direction. Therefore, we can produce all the polarization states of the Poincaré sphere within the beam cross-section.

It is important to note that all skyrmions are Poincaré beams but the converse is not true. The expression $|\Psi_P(r,\phi,z)\rangle)\rangle$ describes that of a skyrmionic beam for $p=0$ and when satisfying the condition $\ell_L \neq \ell_R$.

$$|\Psi_S(r,\phi,z)\rangle = \frac{1}{\sqrt{2}}\left[LG_0^{\ell_R}(r,\phi,z)|R\rangle + exp(i\theta_0)LG_0^{\ell_L}(r,\phi,z)|L\rangle\right]. \tag{4}$$

The polarization distribution generated by a skyrmionic beam under the condition $\ell_L - \ell_R = 1$ with $\ell_L = 1 \& \ell_R = 0$ [$LG_0^{\ell_R}$ mode in right circular polarization (RCP) and $LG_0^{\ell_L}$ mode in left circular polarization (LCP)] is presented in Fig. 4(*a*). The gradient phase difference of $2\pi$ created between the two superposed LG modes results in a rotation of the elliptical polarization with the skyrmion having cylindrical symmetry in its polarization. At (and close to) the center of the beam cross-section, the polarization distribution is right-hand polarized (RHP) and in the peripheral region, is left-hand polarized (LHP) owing to the smaller size of the $LG_0^{\ell_R}$ mode over the $LG_0^{\ell_L}$ mode. The non-uniform polarization distribution produces polarization singularities. The polarization singularities and polarization distribution around them can be experimentally generated via the stokes parameters provided by [35,36]

$$S_0 = \langle E_x E_x^*\rangle + \langle E_y E_y^*\rangle, \tag{5a}$$
$$S_1 = \langle E_x E_x^*\rangle - \langle E_y E_y^*\rangle, \tag{5b}$$
$$S_2 = \langle E_x E_y^*\rangle + \langle E_y E_x^*\rangle, \tag{5c}$$

and

$$S_3 = i\langle E_x E_y^*\rangle - i\langle E_y E_x^*\rangle \tag{5d}$$

with $S_0^2 = S_1^2 + S_2^2 + S_3^2$. At the center of the beam cross-section, we observe perfect CP with an undefined azimuth of polarization, and this results in a C-point singularity ([Fig. 4(*b*) & Fig. 4(*c*)]. In the cylindrical volume formed by the beam cross-section and beam axis (3D space), the polarization singularity is C-line (a 1D singularity, formed at $r = 0$ due to the undefined azimuth of the polarization along $z$), where the Stokes vector components $S_1 = S_2 = 0$. In the transformation from LHP to RHP in the radial direction, we have a linear polarization (LP) along a circular line where the intensity of both the LG modes are the same [Fig. 4(*b*) & Fig. 4(*c*)]. The handedness of the polarization is undefined along this line, and it is referred to as the L-line. In 3D space, it is an L-surface (2D singularity, formed at a particular $r$ due to undefined handedness of polarization along $\phi$ and $z$), where the Stokes vector component $S_3 = 0$. The Stokes vector of the polarization distribution of the beam cross-section is obtained by Eq. 5 and its form is shown in Fig. 4(*d*). The RCP and LCP fields are represented by up and down arrows respectively, while LP is represented by a horizontal arrow. To further understand the relation between the Stokes vector and polarization, the transformation between RHP and LHP Stokes vectors which takes place in the radial direction is illustrated with respect to the polarization in Fig. 4(*e*).

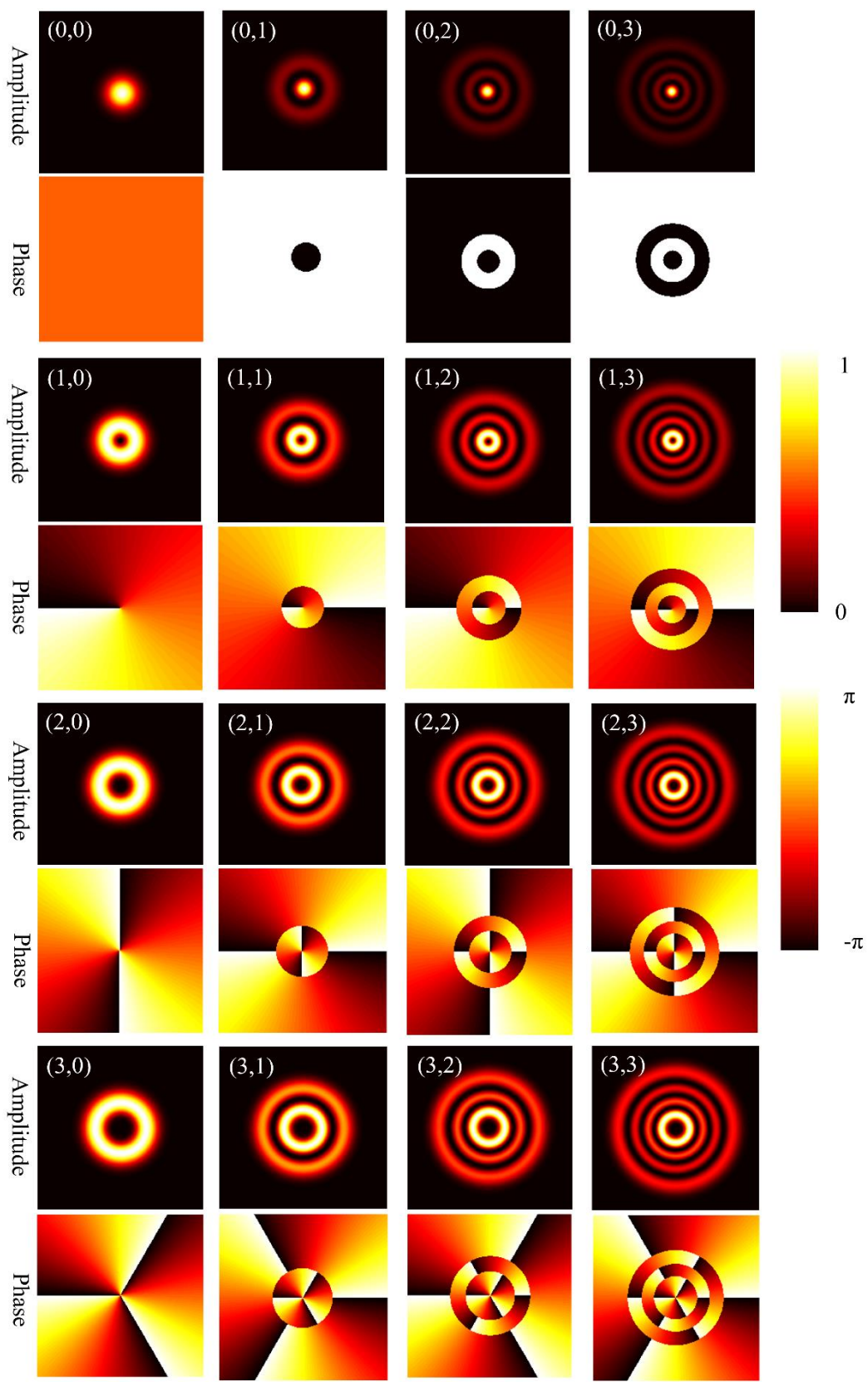

Fig. 3. Images showing the transverse intensity and phase of LG modes with different azimuthal and radial indices ($\ell$, $p$).

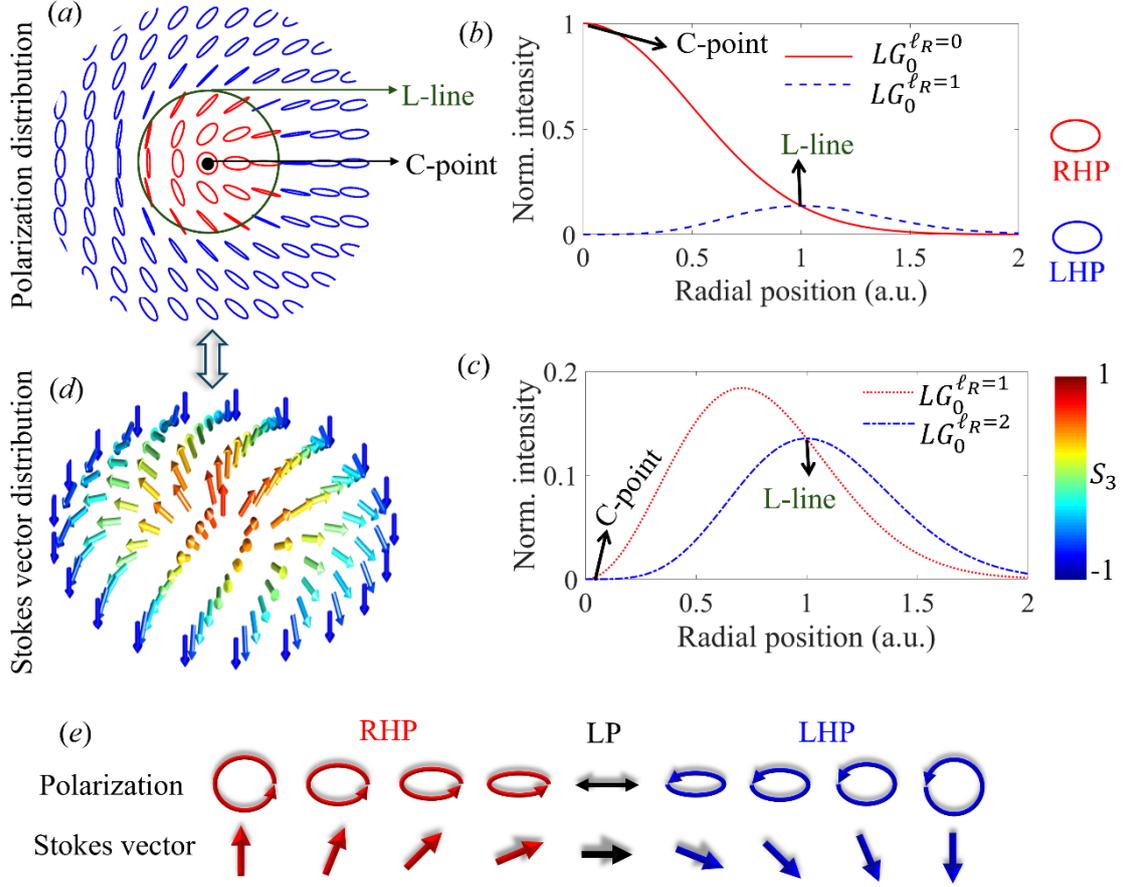

Fig. 4. Images and plots outlining (*a*) the polarization texture of a skyrmion. (*b*) and (*c*) are line profiles of LG modes; (*d*) shows the Stokes vectors in the beam cross-section; and (*e*) shows the relation between polarization and stokes vector, $S_3$. RHP: right-handed polarization, LHP: left-handed polarization.

The Stokes vector of the paraxial skyrmion laser beam obtained by the operation of Pauli matrices σ on the skyrmion state given by Eq.4 can be described as

$$S = \langle \Psi_S(r,\phi,z)|\sigma|\Psi_S(r,\phi,z)\rangle. \tag{6}$$

The $i^{\text{th}}$ component of the associated skyrmion is [5]

$$\tau_i = \frac{1}{2}\epsilon_{ijk}\epsilon_{pqr}S_p\frac{\partial S_q}{\partial x_j}\frac{\partial S_r}{\partial x_k}, \tag{7}$$

here $\epsilon_{ijk}$ is the Levi-Civita symbol with *i*, *j*, *k* ∈ (*x*, *y*, *z*) and *p*, *q*, *r* ∈ (1, 2, 3). The skyrmion field is transverse with no sources or sinks and the associated field lines can only form loops or extend to infinity. The skyrmion number can be obtained by integrating the Stokes field over the entire area of the beam cross-section and it is given in the integral form as [5]

$$N_{sk}(z) = \frac{1}{4\pi}\int \tau_z \, dxdy. \tag{8}$$

Further simplification of the above integral provides the following expression for the skyrmion number

$$N_{sk}(z) = \Delta\ell\left[\frac{1}{1+|v(0,z)|^2} - \frac{1}{1+|v(\infty,z)|^2}\right], \tag{9}$$

where $v(0,z) = exp(i\theta_0)LG_0^{\ell_L}(r,\phi,z)/LG_0^{\ell_R}(r,\phi,z)$. In Eq.9, one of the wave functions $v(0,z)$ and $v(\infty,z)$ is zero and the other is infinity. Therefore, the term in the bracket is either -1 or +1 depending on the superposed LG modes and this determines the sign of the skyrmion number. The magnitude of the skyrmion number is given by $\Delta\ell$ and is merely the order of the skyrmion. The skyrmion number represents the number of times the tip of the vector covers the entire sphere; the skyrmion topology wraps the Poincaré sphere by the number of times given by the skyrmion number. Skyrmionic modes have an integer skyrmion number and they must be conserved throughout their propagation while non-skyrmions have fractional values (for example, $\ell_R = -\ell_L$ in Eq. 4 produce a non-integer skyrmion number).

The three types of skyrmions which are generated in first order Poincaré beams and their stereographic projection on the 2-sphere are illustrated in Fig. 5. The transformation between RHP and LHP takes place in the radial direction, and it is the same for all orders of skyrmions. By examining the polarization and stokes vectors as a function of azimuthal angle, we can determine the type and order of the skyrmion. First-order Poincaré beams have two types of polarization distribution: lemon shape and star shape. The horizontal lemon-shaped polarization distributed Poincaré beam has a Stokes field which exhibits a hedgehog texture in its stokes field and it is considered to be a Néel-type skyrmion.

We can rotate the lemon shape polarization from horizontal to vertical without altering the polarization distribution by providing a phase difference of $\theta_0 = \pi/2$ between the two superposed LG modes. In doing so, the Stokes field distribution changes from having a hedgehog texture to a vortex texture and it becomes a Bloch-type skyrmion. The star shape polarization distribution can be obtained either by interchanging the polarization states by fixing the spatial modes or by exchanging the spatial modes without changing polarization states. The star-shaped Poincaré beam has a saddle texture which belongs to the anti-type skyrmion. As discussed above, the Néel-type and Bloch-type skyrmions can be continuously transformed into each other by changing the relative phase between the superposed modes and therefore they belong to the same topological class. It is worth noting that in this transformation, the skyrmion number is constant. This transformation cannot be done either between Néel-type skyrmion and anti-type skyrmion or between Bloch-type skyrmion and anti-type skyrmion because skyrmions and anti-skyrmions belong to different topological classes.

In the skyrmionic state construction, we have two polarization states and two spatial modes. By permutation and combination, Eq. 4 gives four types of skyrmions. For instance, the skyrmion number of arbitrary skyrmion order generated by the superposition of Gaussian ($LG_0^0$) and arbitrary order Gaussian vortex ($LG_0^\ell$) modes for different permutations and combinations of ($\ell_R$, $\ell_L$) are tabulated in Table 2. The skyrmions can have a skyrmion number of either -$\ell$ or +$\ell$. This permutation and combination as applied to first order skyrmions yields six types of skyrmions with skyrmion number $N_{sk} = \pm 1$ (Fig. 6). An arbitrary skyrmion order can also be generated for the combination of ($\ell_R \neq 0$, $\ell_L \neq 0$).

Table 2. Skyrmion number of skyrmions and anti-skyrmions in terms of winding number or topological charge of the LG modes involved in skyrmion generation.

| $\ell_R$ | $\ell_L$ | $N_{sk}$ | Skyrmion type |
|---|---|---|---|
| 0 | +$\ell$ | -$\ell$ | Néel-type and Bloch-type |
| -$\ell$ | 0 | +$\ell$ | |
| 0 | -$\ell$ | +$\ell$ | anti-type |
| +$\ell$ | 0 | -$\ell$ | |

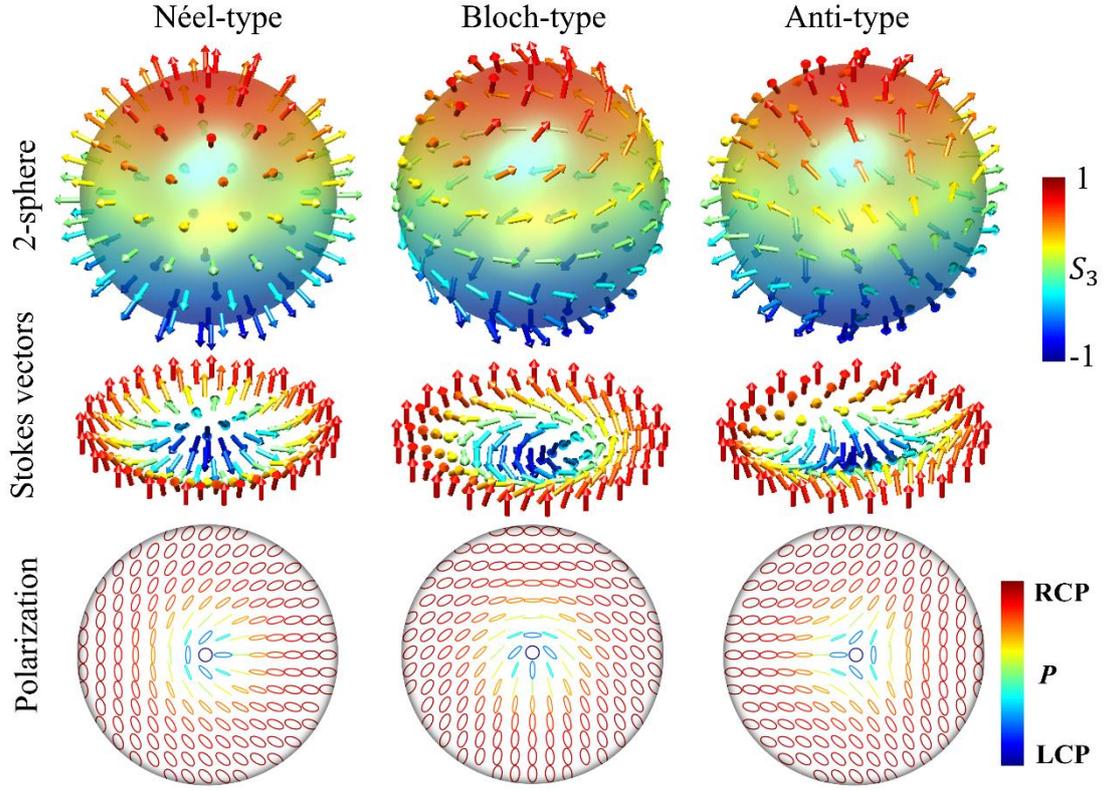

Fig. 5. Images depicting the stereographic projection of first-order skyrmionic quasiparticles created in the beam cross-section onto a 2-sphere.

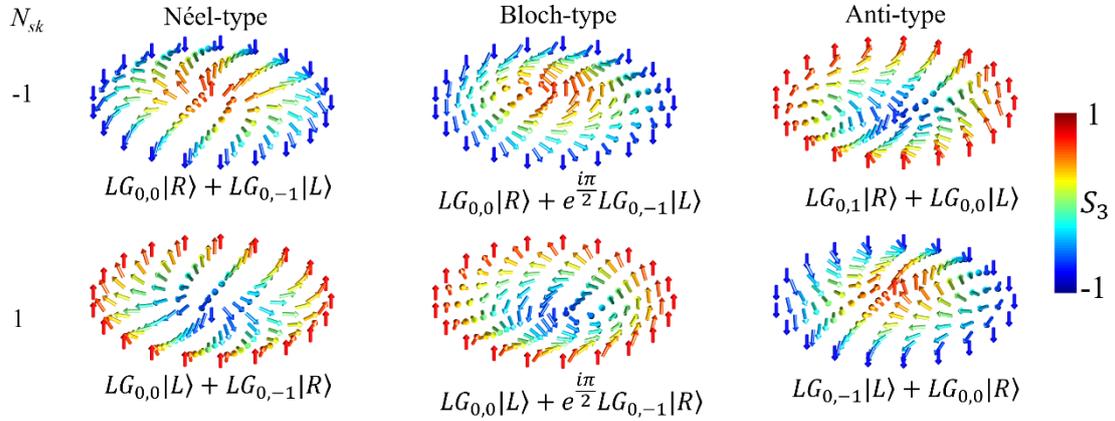

Fig. 6. Images depicting first order skyrmions and anti-skyrmions with skyrmion number $N_{sk}$ = -1 (first row) and $N_{sk}$ =1 (second row).

The skyrmions created in paraxial laser beams are obtained by the superposition of two orthogonal LG modes; these orthogonal LG modes have different Gouy phase. The Gouy phase [37-39] difference between the two superposed modes produces a rotation of the polarization pattern as a function of propagation distance, however, the skyrmion number remains unchanged. The propagation of a first-order skyrmion under a focusing condition is presented in Fig. 7. We can clearly see the evolution of its polarization pattern and the transformation of the skyrmionic texture between Néel-type and Bloch-type skyrmions without changing the skyrmion number value (+1) at every transverse plane. On the 2-sphere, the Stokes vectors at the north and south poles are fixed and all other vectors effectively rotate with propagation. This scenario is true for all optical quasiparticles generated in paraxial laser beams.

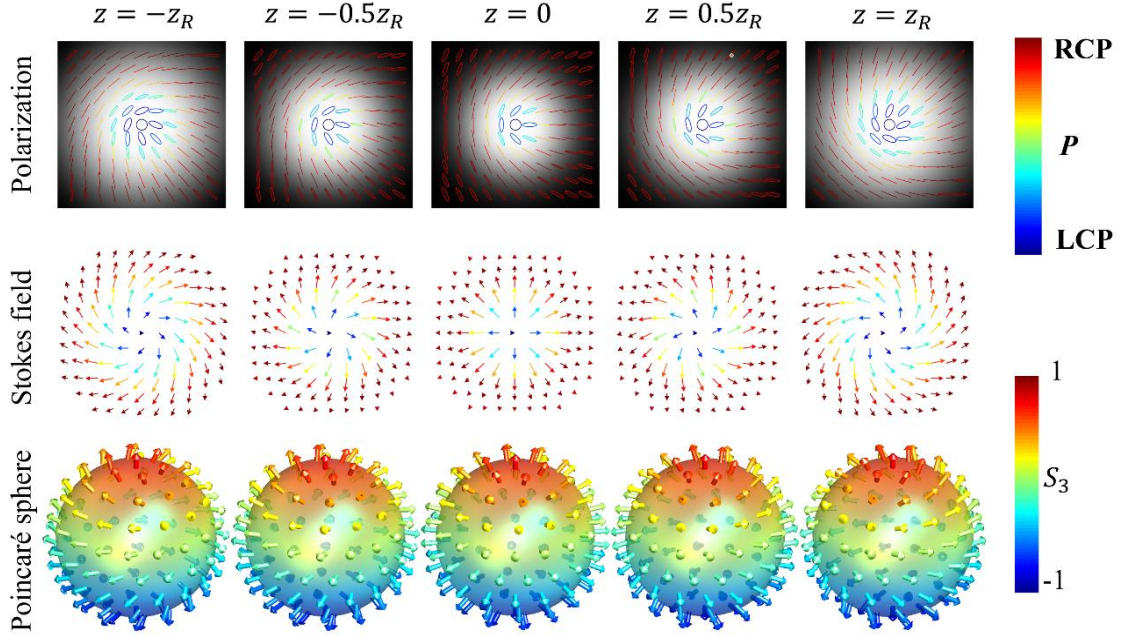

Fig. 7. Images showing the propagation dynamics of a first-order skyrmion under focusing conditions and various axial positions z. The top row depicts the polarization distribution of the skyrmion plotted over its intensity distribution; the middle row depicts the stokes field of the skyrmion in its beam cross-section; and the bottom row depicts the projection of the Stokes field of the skyrmion onto a Poincaré sphere.

Further understanding of skyrmionic textures in paraxial vector beams can be gleaned through the investigation of rotation of non-uniform polarization distributions. The rotation of the Stokes vector and polarization axis in the azimuthal direction is governed by the difference in OAM number $\Delta\ell$. $\Delta\ell$ in the skyrmion rotates the Poincaré vector $\Delta\ell$ times on a path enclosing the vortex. The corresponding polarization ellipse rotates by only half the amount. For reference, we plotted the polarization rotation of skyrmions up to order 3 in Fig. 8. For skyrmions of order 1, 2 and 3, the respective rotation of the Poincaré vector is 1, 2, and 3 times, while the respective rotation of the polarization is 1/2, 1, and 3/2 times. The skyrmions also have rotational symmetries in their polarization textures. The Néel-type and Bloch-type skyrmions have one-fold symmetry in their 1st, and 3rd orders and full symmetry in their 2nd order. The anti-skyrmions have three-fold, four-fold, and five-fold symmetries in the respective 1st, 2nd, and 3rd orders. Rotation of the polarization axis of skyrmions is anti-clockwise while anti-skyrmions exhibit clockwise rotation. We can clearly observe this scenario in experiments through polarization analysis. When the skyrmion passes through the polarization analyzer, we observe a petal structure that rotates in the direction of the polarizer rotation; however, this is reversed in the case of anti-skyrmions.

Skyrmions can also be classified into bright and dark skyrmions based on their intensity distribution. When one of the azimuthal indices in Eq. 4 is zero, skyrmionic beams have a bright central region with a super-Gaussian distribution and are called bright skyrmions. In the case of skyrmionic beams with non-zero azimuthal indices, they exhibit the dark core of both superposed modes and a characteristic doughnut shape intensity distribution; these are referred to as dark skyrmionic beams. In dark skyrmions, the smaller azimuthal index mode dominates at the center. In Fig. 9, we depict all three types of skyrmions with skyrmion number $N_{sk} = \pm 1$ in both bright and dark form. For a fixed Gaussian beam size, the size of a skyrmion increases with its order. The bright skyrmions with the same order have the same size but this is not true in the case of dark skyrmions. The size of the dark skyrmion of a given order increases with the

OAM number of its constituent LG modes. For example, first-order skyrmions formed by ($\ell_R$, $\ell_L$) = (1,2), (2,3), (3,4) have different sizes, with the increase in their size in ascending order.

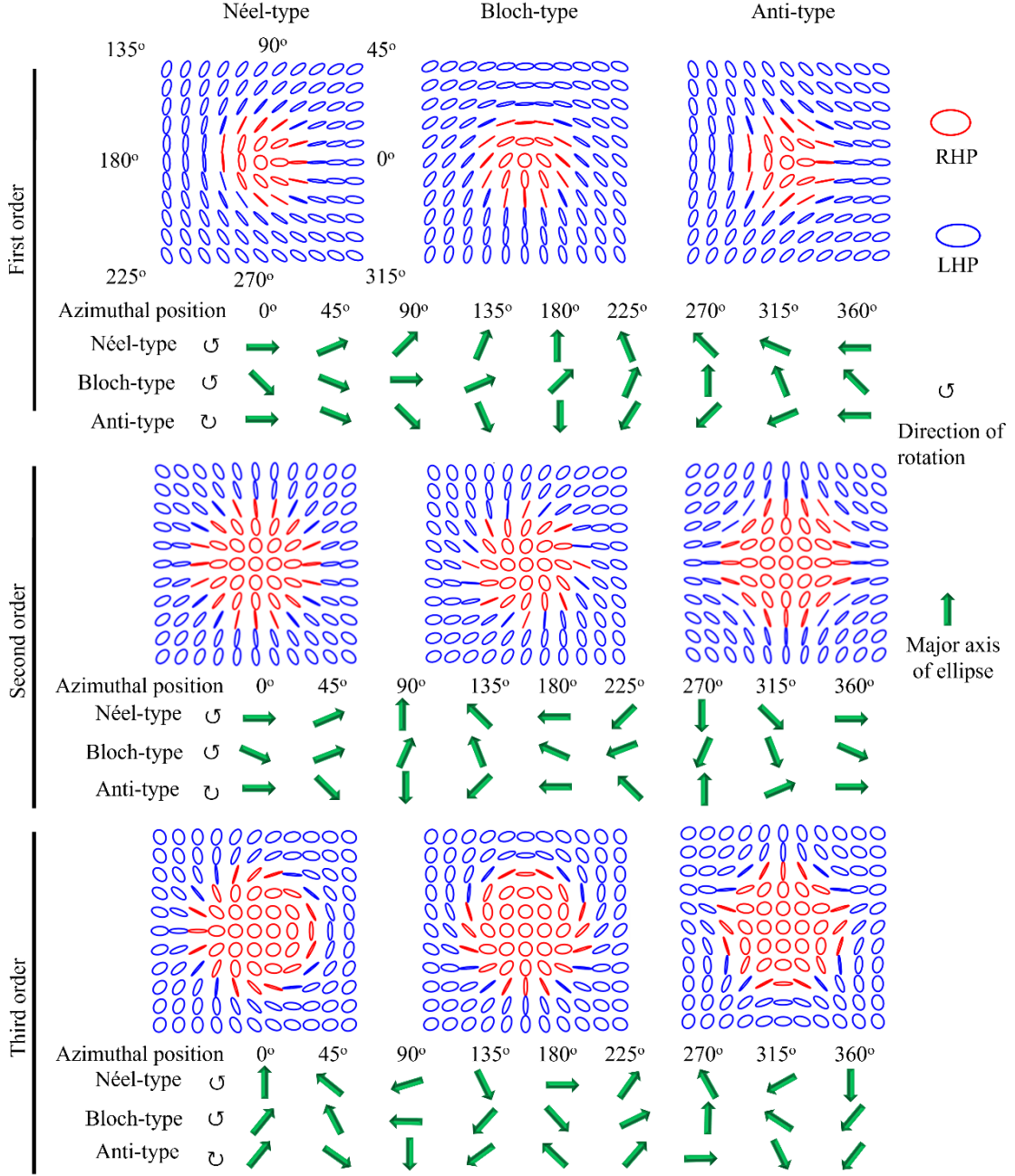

Fig. 8. The rotation of polarization in the azimuthal direction for the first three orders of skyrmions. The major axis of polarization is a director but here we represent it with a single arrow to visualize its rotation.

**2.2. Skyrmions in paraxial Bessel beams**

Bessel beams are another cylindrical symmetric solution of the paraxial wave equation and have order-dependent azimuthal phase and radial intensity distribution [40-42]. Bessel beams also have unique properties, exhibiting non-diffraction and a self-healing nature. We can obtain all the skyrmion textures in Bessel beams similarly to LG beams. By replacing the LG modes with Bessel modes in Eq.4, we can obtain skyrmions as follows

$$|\Phi_s(r,z)\rangle = \frac{1}{\sqrt{2}}\big[B_{\ell_R}(r,\phi,z)|R\rangle + exp(i\theta_0)B_{\ell_L}(r,\phi,z)|L\rangle\big]. \tag{10}$$

Here, $B_\ell$ is the $\ell^{th}$ order ideal Bessel beam and is given by

$$B_\ell(r,\phi,z) = exp(ik_z z - i\omega t) J_\ell(k_r r) exp(i\ell\phi). \tag{11}$$

Here, $k_z$ and $k_r$ are the respective longitudinal and radial components of propagation vector $k$ and are related by $k = \sqrt{k_r^2 + k_z^2}$. The angular frequency of the Bessel beam is $\omega$. Also, $J_\ell(.)$ is the Bessel function of the first kind. The Bessel beam provided by Eq. 11 is ideal and it has an infinite spatial extent and preserves its shape for an infinite propagation distance. Thus, Eq. 10 describes a Bessel skyrmion with an infinite propagation distance. Such a beam cannot be realized in practice as it is not possible to produce an infinitely-extended Bessel beam in experimental laboratories. Bessel beams are typically generated in experimental setups via wavefront division interference [43-45]. These beams exhibit a Bessel-like nature over a finite range and are often called quasi-Bessel beams. Given this, we can only generate quasi-skyrmions using experimentally produced Bessel beams. To achieve Bessel mode conversion, we must provide radial and azimuthal phase variation to a Gaussian mode. One of the simplest methods of generating a Bessel beam with an arbitrary order $\ell$ is by pumping an axicon with a LG mode with azimuthal index $\ell$ and radial index $p = 0$ (Eq. 2). The intensity distribution of the Bessel beam formed by pumping the axicon with a Gaussian vortex mode can be mathematically expressed as [46]

$$B_\ell^2(z,r) \propto \left(\frac{z}{z_{max}^G}\right)^{2\ell+1} exp\left(\frac{-2z^2}{{z_{max}^G}^2}\right) J_\ell^2(k_r r). \tag{12}$$

Here, $k_r = sin[(n-1)\alpha]$ is the radial component of the propagation vector and $z_{max}^G = kw/k_r$ is the range of the Bessel beam formed in the presence of a Gaussian beam as a pump source, with $\alpha$ and $n$ as the base angle and refractive index of the axicon.

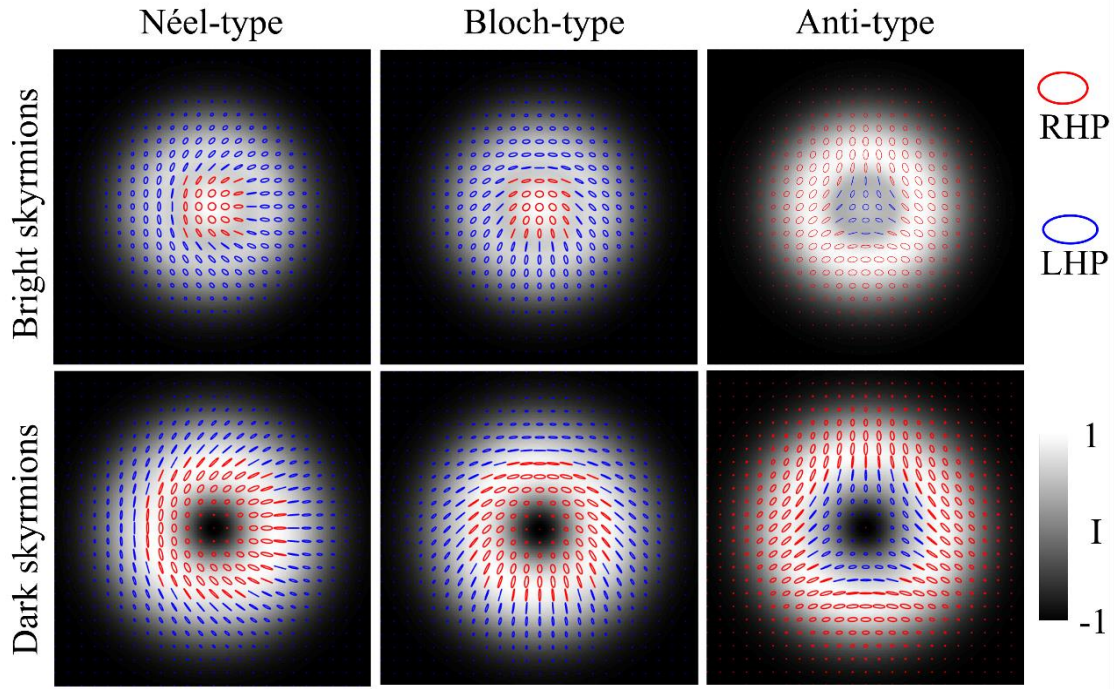

Fig. 9. Polarization distribution of first-order skyrmions. First row: bright skyrmionic beams ($\ell_R$, $\ell_L$) ∈ (0, 1); and second row: dark skyrmionic beams ($\ell_R$, $\ell_L$) ∈ (0, 1).

It is important to note that the inner and outer radii of the Gaussian vortex mode, represented with $r_1$ and $r_2$ respectively, constrain the longitudinal intensity distribution of the Bessel beam. Indeed, the onset and offset positions of the Bessel beam of order $\ell$ can be written in terms of its pump Gaussian vortex radii as $z^\ell_{min} = r_1/(n-1)\alpha$, and $z^\ell_{max} = r_2/(n-1)\alpha$ [47]. Also, the range and peak position of the Bessel beam along its propagation direction are given by $z^\ell_{range} = (r_2 - r_1)/(n-1)\alpha$ and $z^\ell(I_{max}) = z^G_{max}(2\ell+1)^{1/2}/2$ respectively. The longitudinal intensity distribution of Bessel beams is thus order dependent [see Fig. 10]. For effective generation of stable skyrmionic textures in Bessel beams, the transverse and longitudinal modes of superposed Bessel modes must be matched. Thus, the major challenge facing the generation of Bessel skyrmions (i.e., propagation independent Poincaré beam) is longitudinal mode matching.

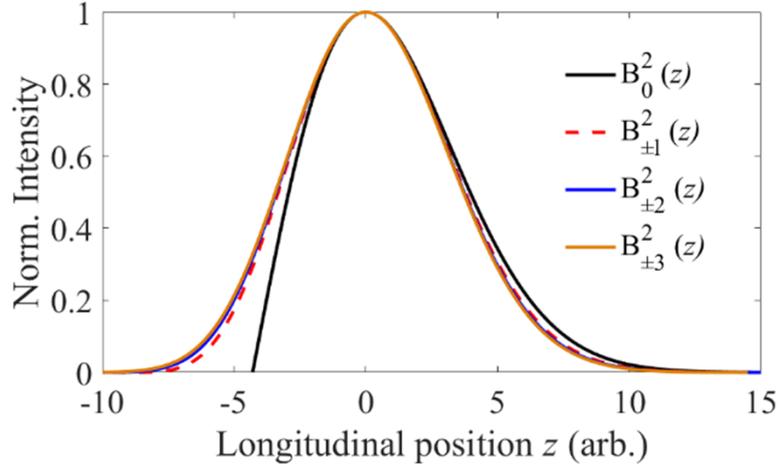

Fig. 10. Plot showing the line profiles of longitudinal mode matched Bessel beams at their peak intensity (the intensities are normalized with the peak intensity of their first maxima).

Three research groups have successfully demonstrated Poincaré modes using Bessel beams. The first demonstration was by V. Shvedov *et. al.* [48] wherein Poincaré modes were produced by perfectly matching the 0$^{th}$ and 1$^{st}$ order Bessel modes using conical diffraction. The Bessel beams were formed using a high optical quality biaxial crystal [49]. Subsequent demonstrations by the other two research groups utilized superposition of two Bessel beams with different radial propagation vectors in order to generate a Poincaré beam [50,51].

In 2022, K. Singh *et. al.* [52] demonstrated the generation of quasi-skyrmions by superposing two Bessel Gaussian modes with different propagation vectors. In this work they also developed a theory that is similar to the theory used to support LG based skyrmionic modes. Based on their theory the skyrmion number of Bessel skyrmions is provided by

$$N_{sk} = q\Delta\ell \left[ \frac{1}{1+\delta(0)^2} - \frac{1}{1+\delta(r_0)^2} \right] \qquad (13)$$

with $\delta(r) = \left| B_{\ell_L}(rk_r)/B_{\ell_R}(rk_r) \right|^q$, $\Delta\ell = \ell_R - \ell_L$, and $q = (|\ell_L| - |\ell_R|)/||\ell_L| - |\ell_R||$. The skyrmion number obtained by using Eq. 13 is in the localized circular region bounded by the condition $r \leq r_0$ and the sign and amplitude of the skyrmion number are governed by $q\Delta\ell$. In Bessel skyrmions, periodic changes in polarization between RHP and LHP occurs along the radial direction and is attributed to Bessel rings [Fig. 11(*a*) and Fig. 11(*b*)]. At the center of the beam, a C-point singularity exists with CP. The continuous transformation between RHP and LHP results in L-line singularities where the intensities of superposed Bessel modes are the same. As shown in Fig. 11(*c*), we can draw the concentric circles around the beam axis with their radii following the periodic change in the handedness of the polarization. The estimated skyrmion number in the central region is 1 and the skyrmion number in concentric annulus regions around

the beam axis evaluates to 0 [52]. Skyrmionic textures in Bessel profiles can be produced not only for $k_r^L = k_r^R$ but also for $k_r^L \neq k_r^R$ and the first zeros of the Bessel beams do not overlap. In the case of $k_r^L = k_r^R$, Bessel skyrmions do not undergo transformation between Néel-type and Bloch-type skyrmions owing to zero Gouy phase difference, which is in contrast to what is observed in LG skyrmions. However, for $k_r^L \neq k_r^R$ we encounter the Gouy phase difference between the two superposed Bessel modes which vary linearly with propagation. Therefore, in this case, we observe periodic transformation between Néel-type and Bloch-type skyrmions [Fig. 11(d)].

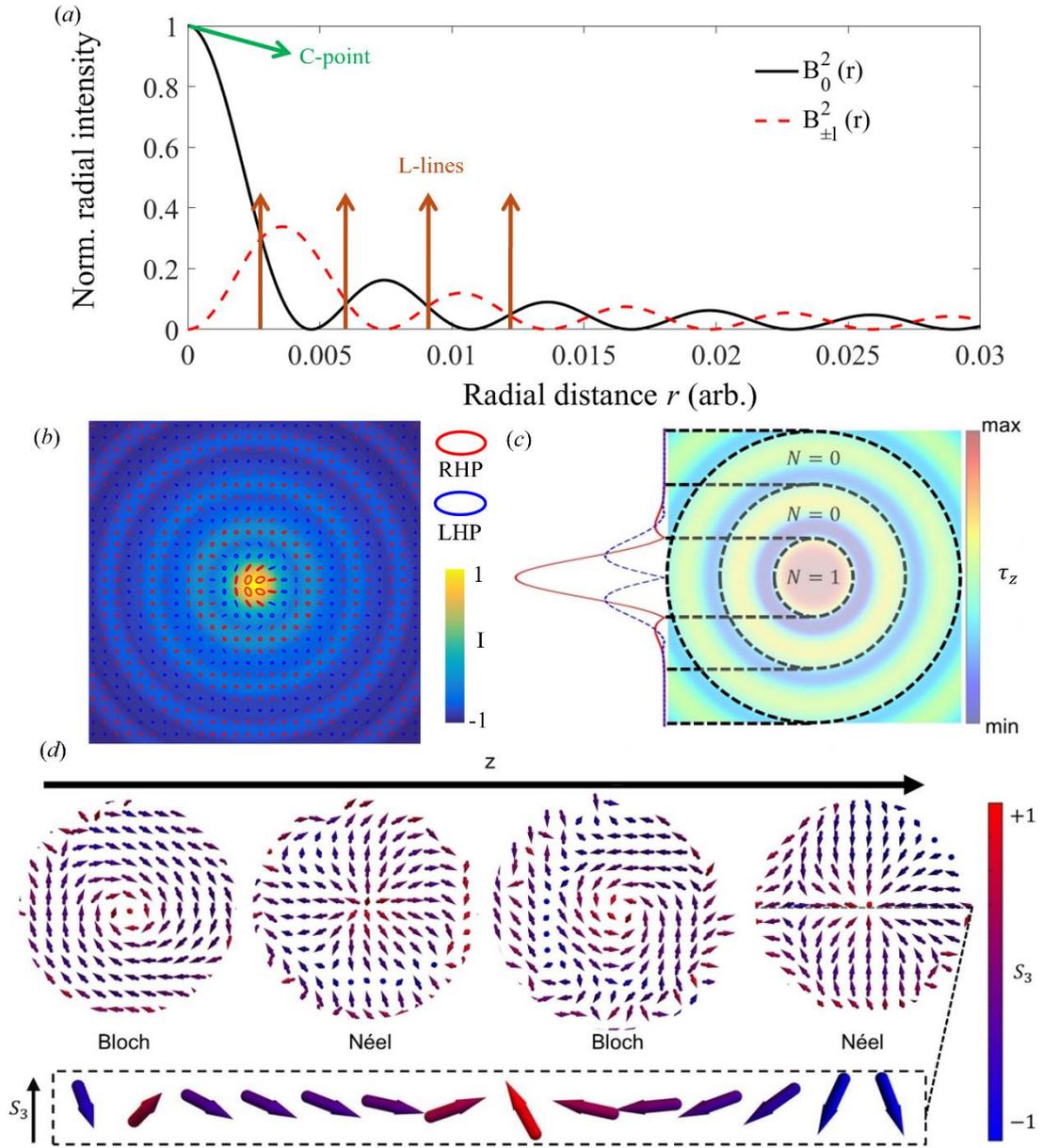

Fig. 11. Images highlighting (a) the transverse line profiles of $0^{th}$ and $1^{st}$ order Bessel beams. (b) An overlay of the polarization on top of the intensity profile of a first order Néel-type Bessel skyrmion. (c) The skyrmion number distribution across the beam cross-section. (d) The Stokes vector distribution in the central region of an experimentally generated Bessel skyrmion, as a function of propagation (z). The mismatch in the $k_r$ of superimposed Bessel beams produced propagation-dependent transformation between Néel-type and Bloch-type skyrmions [the Stokes vectors plotted here are located in the central circular region of Fig. 11(c)]. [Fig. 11(c) and Fig. 11(d) are adapted from [52].

Recently, we reported a simple and robust technique for the generation of high-power Bessel skyrmions in continuous wave (cw) and pulsed laser sources [53]. This technique is based on longitudinal mode matching of Bessel modes. This can be achieved by properly tuning the transverse profile of the initial beam used in Bessel mode conversion.

An LG mode with azimuthal index $\ell$ and radial index $p = 0$ has a doughnut shape intensity profile with an azimuthal phase (Eq. 2). We can also consider a beam which has a similar doughnut shape intensity profile which does not have a helical wavefront, called a Hollow Gaussian (HoG) beam, which can be mathematically expressed as [54]

$$E_m(r) = \sqrt{\frac{2^{2m+1}P}{(2m)!\,\pi w^2}} \left(\frac{r^2}{w^2}\right)^m exp\left(\frac{-r^2}{w^2}\right), \tag{14}$$

here, $m$ is the order of the HoG beam. This HoG beam can be easily produced experimentally using techniques utilizing SLM, SPP, and thermal lenses [55-58]. From Eq. 2 and Eq. 14, we can clearly observe that a LG mode with azimuthal index $\ell$ and radial index $p = 0$ and a HoG with order $m = \ell/2$ have the same transverse intensity distributions. Hence, the Bessel beams generated from these two beams have the same longitudinal intensity distribution. The $0^{th}$ order Bessel beam produced by HoG of order $m$ is given by [59]

$$B_0^2(z,r) \propto \left(\frac{z}{z_{max}^G}\right)^{4m+1} exp\left(\frac{-2z^2}{z_{max}^{G\,2}}\right) J_0^2(k_r r). \tag{15}$$

As shown in Fig. 12, we can achieve perfect longitudinal mode matching between $0^{th}$ and $\ell^{th}$ order Bessel beams by utilizing an axicon with both LG and HoG modes satisfying the $m = \ell/2$ condition. The polarization distribution of first-order Néel-type, Bloch-type, and anti-type Bessel skyrmions generated based on this technique are depicted in Fig. 13. The skyrmionic textures fabricated by this technique preserve their shape within the Bessel range without any diffraction and they become non-diffracting optical skyrmions. Also, these skyrmions acquire a self-healing nature from their parent Bessel beams as an inherited property and can preserve their skyrmionic textures even after encountering obstacles [51,53,60,61]. In this technique, the two superposed Bessel modes have the same propagation vector and as a result, there is no transformation between Néel-type and Bloch-type skyrmion textures as they propagate. We can also generate dark skyrmions in the Bessel profile by superposition of two higher-order Bessel beams, however, due to longitudinal mode mismatch, the production of stable skyrmionic textures takes place in over a limited propagation range. This limited propagation range can be increased by increasing the order of both of the superposed Bessel beams.

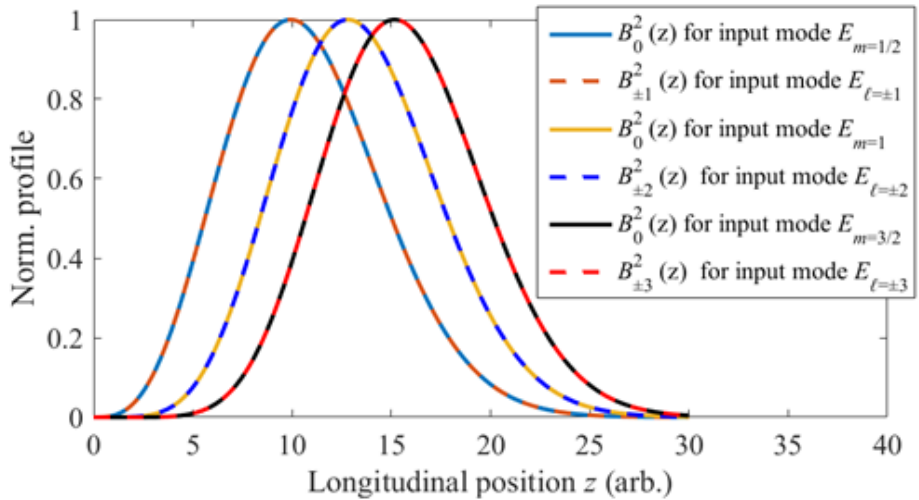

Fig. 12. Plot of the longitudinal line profiles of Bessel beams generated by using Laguerre Gaussian modes with azimuthal index $\ell$ and radial index $p = 0$ and hollow Gaussian modes as modes incident onto an axicon (reproduced from [53]).

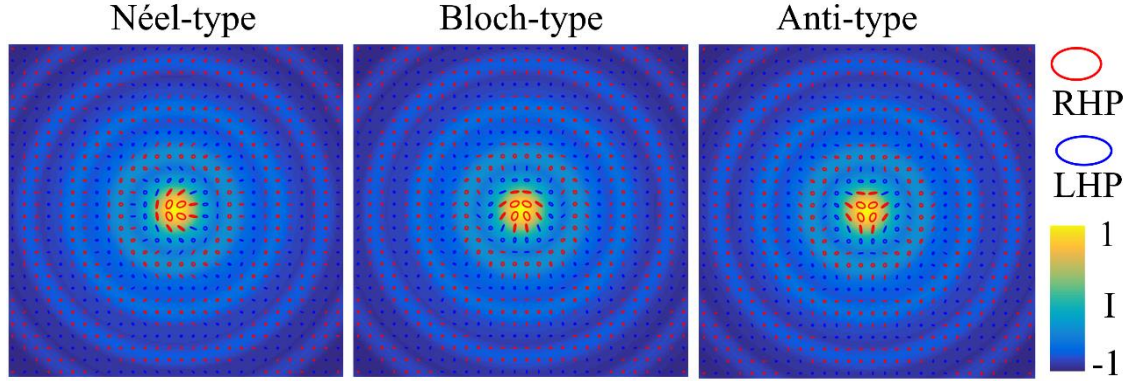

Fig. 13. The transverse polarization distribution of Bessel skyrmions generated via perfect longitudinal mode matching of a Laguerre Gaussian mode with azimuthal index $\ell$ and radial index $p = 0$ and a hollow Gaussian mode with order $m = \ell/2$, both as beams incident onto an axicon.

### 3. Bimerons

Another interesting quasiparticle is a meron which is a half skyrmion and confines its texture on hemi 2-sphere [62-65]. Merons can also have higher-order topologies like skyrmions. In paraxial merons, the gradient polarization distribution takes place as a function of azimuthal and radial coordinates in the beam cross-section. In the radially outward direction, the transformation of either RCP/LCP to LP takes place [Fig. 14]. Merons have a C-point at the center with the LP boundary. Similar to skyrmions, all merons have only one C-point and their topological texture bounded by an L-line. These particles are single-hand elliptical polarization vector beams.

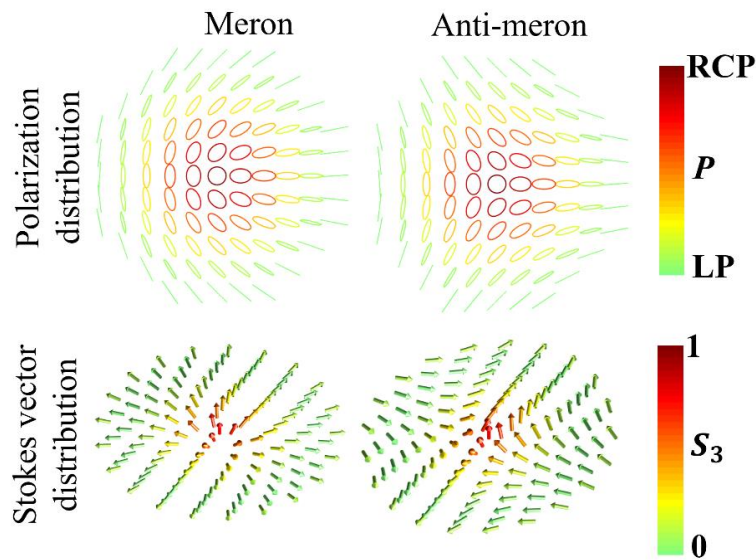

Fig. 14. Images showing the textures of meron and anti-meron quasiparticles in paraxial laser beams.

The generation of a meron in paraxial laser beams has yet to be demonstrated experimentally (as they are inherently very difficult to generate), however meron lattice structures can be generated experimentally.

Meron lattices in paraxial laser beams can be generated by replacing the CP basis with the LP basis in the superposition state formed by LG modes (Eq. 4). The superposition state becomes

$$|\Psi_M(r,\phi,z)\rangle = \tfrac{1}{\sqrt{2}}\left[LG_0^{\ell_R}(r,\phi,z)|H\rangle + exp(i\theta_0)LG_0^{\ell_L}(r,\phi,z)|V\rangle\right], \tag{16}$$

where, $|H\rangle$ and $|V\rangle$ are the horizontal and vertical polarizations. The meron lattices generated by this method are often called bimerons [66,67]. For $\Delta\ell = \pm 1$, Eq. 16 yields a first order bimeron with two merons containing opposite circular polarizations (Fig. 15). The number of merons in the bimeron increases with increasing $\Delta\ell$ in even numbers. For $\Delta\ell = \pm 2$, Eq. 16 yields a second order bimeron with four merons. Each meron has a C-point singularity and is separated from its neighbor by an L-line. These meron lattices have a petal structured polarization distribution. The number of merons in the lattice can be increased in the azimuthal direction and confined within the radial coordinate by increasing $\Delta\ell$. These textures are topologically protected and propagation invariant. Moreover, rotation of their texture occurs due to the Gouy phase difference created between the superposed modes, but their topology is preserved.

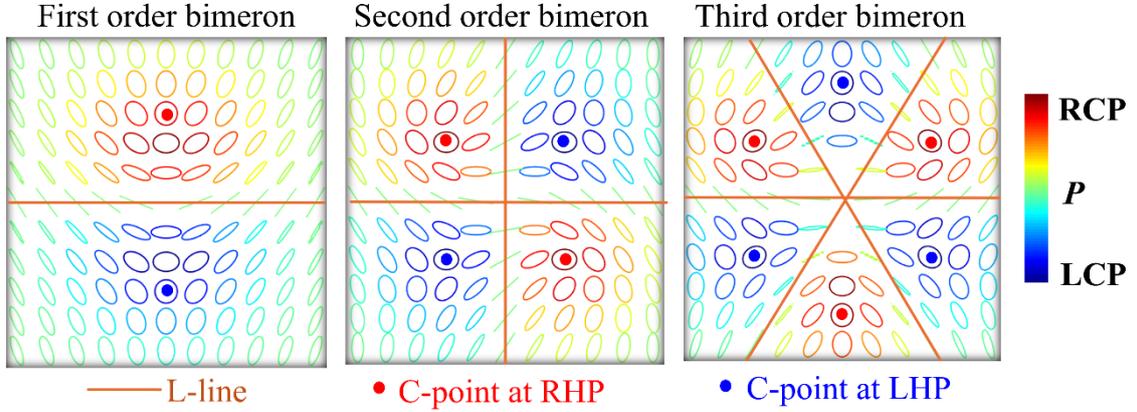

Fig. 15. Polarization texture of bimerons created in paraxial laser beams. A first order bimeron was generated for $\Delta\ell = \pm 1$ (it possessing two merons) and a second order bimeron was generated for $\Delta\ell = \pm 2$ (it possessing four merons).

In a similar fashion to skyrmions, bimerons can also be stereographically projected onto a 2-sphere whose north and south poles are in the LP basis. As shown in Fig.16, these modes can also exhibit Néel-type, Bloch-type, and anti-type textures. The azimuthal rotation of the polarization texture in the beam cross-section as a function of propagation due to the Gouy phase created under focusing conditions results in rotation of the 2-sphere around its polar axis.

From Fig. 5 and Fig. 16, we can see that skyrmions and bimerons can be stereographically mapped onto a 2-sphere which is formed by respective circular and linear polarization bases. The CP basis can transform into a LP basis, and the vice versa is also true. By this transformation, we can also transform the polarization texture between skyrmions and bimerons without changing the topology, a process termed homeomorphism. Indeed, we can project skyrmions and bimerons generated from the same LG modes onto a single 2-sphere constructed either by a LP basis or a CP basis. For example, in Fig. 17, we show the projection of a first-order skyrmion and bimeron onto a single 2-sphere formed using a CP basis. Here, the skyrmion state is projected from one pole and ends at the other pole, while the bimeron state is projected from a side point and ends at its antipodal point on the 2-sphere. For simplification, here we construct bimerons in a LP basis, however, it is not limited to it. For example, if we replace the LP basis with an arbitrary pair of orthogonal polarization states, the obtained bimeron projection starts from the corresponding pair of antipodal points on the parametric sphere.

In addition to single skyrmionic beams, we can also fabricate skyrmion lattices in paraxial laser beams by using a suitable combination of structured laser beams [68]. The skyrmion number of individual quasiparticles present in a single optical beam can be independently obtained and characterized by techniques like rational maps and contour integration [69].

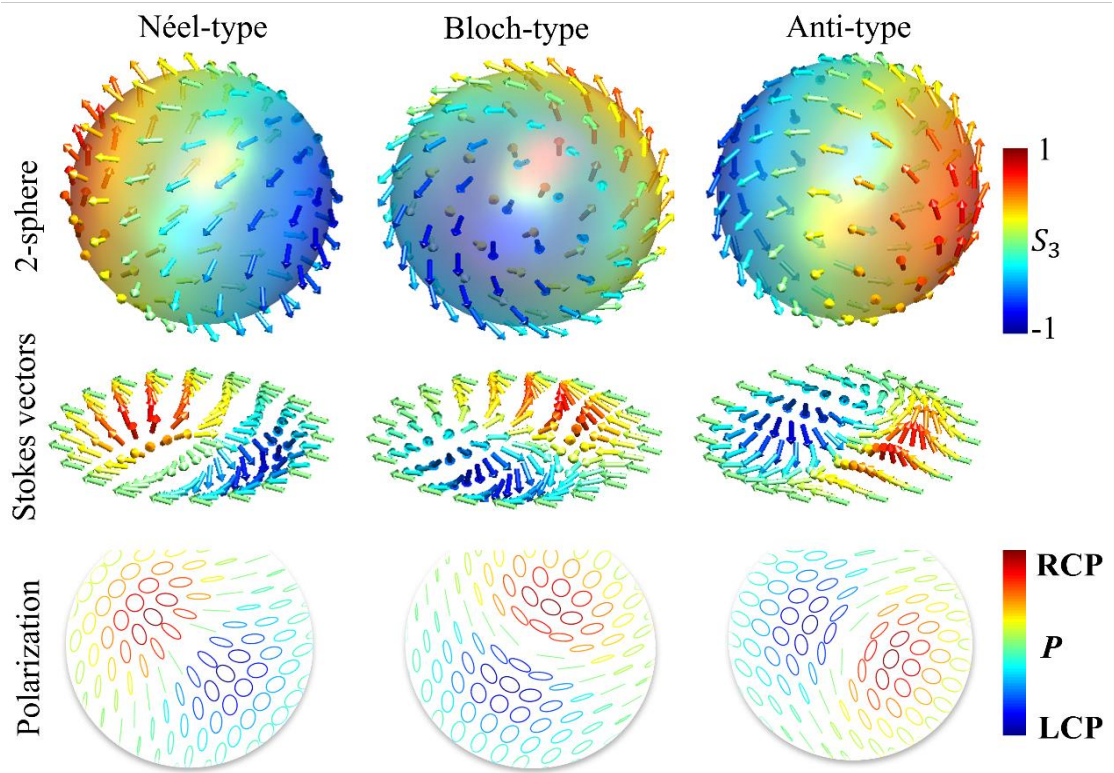

Fig. 16. Images detailing the stereographic projection of bimerons onto a 2-sphere.

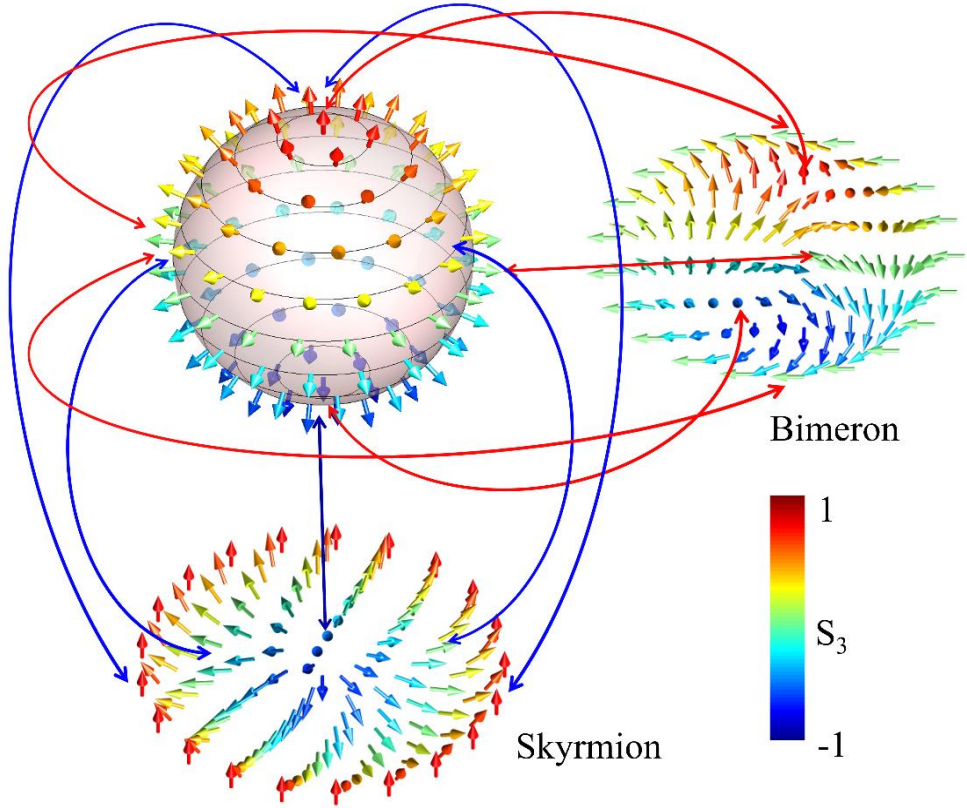

Fig. 17. Stereographic projection of a bimeron and a skyrmion from a single 2-sphere (reproduced from [67]).

It is a straightforward process to generate non-diffracting and self-healing bimerons in Bessel beams by simply replacing LG modes with Bessel modes in Eq. 16.

## 4. Skyrmioniums and Bimeroniums

In the preceding sections, the skyrmionic and bimeronic textures were generated based on LG modes with radial index $p = 0$. In these textures, the handedness of polarization transforms either from LCP to RCP or from RCP to LCP. We can also generate more complex optical quasiparticles [70] in paraxial laser beams by using Eq. 3 and considering a non-zero radial index $p \neq 0$ and $\ell_L \neq \ell_R$. The textures created in paraxial laser beams based on Eq. 3 in a CP basis are called skyrmioniums [71] and in a LP basis are called bimeroniums [72] [Fig. 18]. For a given order, these textures can also be classified into Néel-type, Bloch-type, and anti-type textures. Skyrmioniums and bimeroniums have radially dependent periodic variations in the handedness of polarization. The number of transformations between the orthogonal handedness of polarizations in the radial direction depends on the radial index value. In the present context, we consider the case where $p = 1$.

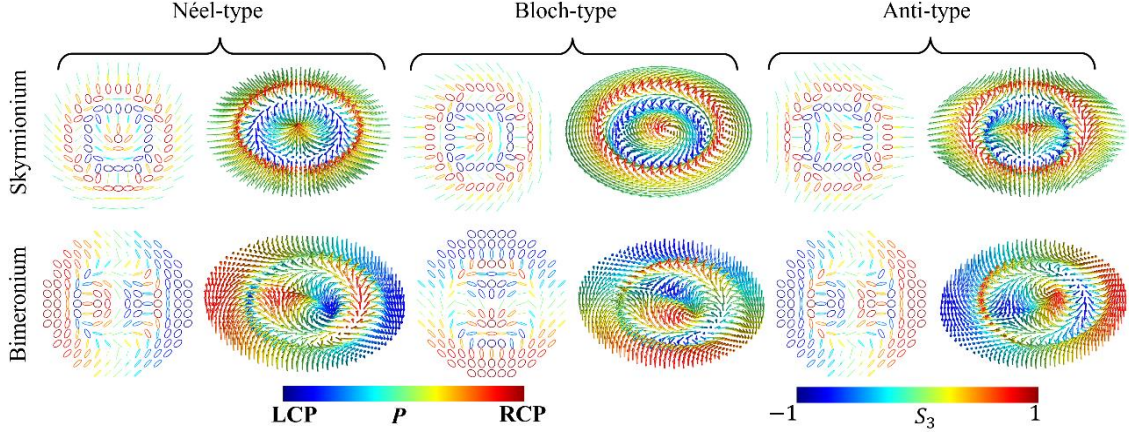

Fig. 18. Images detailing Néel-type, Bloch-type, and anti-type skyrmionium and bimeronium textures created in paraxial vector beams.

## 5. Skyrmionic hopfions

The particle-like continuous fields in 3D space called skyrmions and 3D topological solitons (called hopfions), have garnered significant interest across multiple research fields due to their unique nontrivial 3D topological textures. These textures have been experimentally generated in liquid crystals [73] and cold quantum physics experiments [74]. Even though these particles have complex structures, we can easily understand them by projecting their properties/characteristics from real space onto a spherical volume [75-77]. Skyrmionic textures can be projected from 3D space onto a hypersphere in 4D space (3-sphere) through homomorphism. Similarly, hopfions can be represented on a 2-sphere constructed in 3D space.

In the previous sections of this review, we have detailed and discussed the properties of optical quasiparticles which exist as 2D topological states and are confined in a plane. Similarly, we can also construct 3D topological states called skyrmionic hopfions in structured laser beams, under a paraxial approximation. Recently, D. Sugic, *et. al.* theoretically proposed and experimentally demonstrated skyrmion and hopfion textures by describing them in terms of polarization and phase within the 3D volume of paraxial structured laser beams (referred to as a skyrmionic hopfion beam) [78]. Here, the skyrmions wrap their configuration around the hypersphere an integer number of times and hence they can be quantified with integer topological numbers termed the skyrmion number. Their hopfion nature comes from the linking of projected fibers (these fibers in effect represent polarization states and are discussed in greater detail below). Thereby, the skyrmion number is the linking number of each of the fibers. The hopf map of skyrmionic hopfions can be mathematically expressed as

$$h(u,v) = [2\text{Re}(u^*v), 2\text{Im}(u^*v), |u|^2 - |v|^2], \tag{17}$$

and $u, v \in \mathbb{C}$ with $|u|^2 + |v|^2 = 1$. Here, the map from 3-sphere to 2-sphere, $h: S^3 \to S^2$ defines the mathematical hopf fibration. The generalization of this map is in the form of $h(u^m, v^n)$ with integers $m$ and $n$ and $|u^m|^2 + |v^n|^2 = 1$. The fibration depends on these indices and if they are co-primed then the fibers will be knotted. The necessary wave functions for building the optical skyrmionic hopfions in the vector laser beams are derived from two light fields which are orthogonally circularly polarized, i.e., $(u, v) = (E_R, E_L)$ and

$$h(E_R, E_L) = [2\text{Re}(E_R^*E_L), 2\text{Im}(E_R^*E_L), |E_R|^2 - |E_L|^2]. \tag{18}$$

It should be noted that utmost care must be taken when selecting suitable laser modes in the superposition, in order to achieve high-quality optical skyrmionic hopfions in the light fields. The general state of the optical skyrmionic hopfion in the laser modes can be expressed in terms of the superposition of orthogonally polarized light fields as

$$|\Psi_{sh}(r,\phi,z)\rangle = E_R(r,\phi,z)|R\rangle + E_L(r,\phi,z)|L\rangle, \tag{19}$$

here, $E_R(r,\phi,z)$ and $E_L(r,\phi,z)$ are the wave functions of two light fields in the respective LCP and RCP states. These two scalar light fields are structured in a suitable way to realize the 3D topological structure of skyrmionic hopfions in laser modes. In skyrmionic hopfions, Neel-type, Block-type, anti-type, and intermediate textures can also be observed, and we can transform between these textures by tuning the relative phase and interchanging the polarization states between the two structured modes [79].

Cylindrically symmetric 3D polarization distributions as required for the generation of hopfion structures in light fields can be easily realized by utilizing LG modes (Eq.2) as the primary modes in the superposition. The scalar modes $E_R(r,\phi,z)$ and $E_L(r,\phi,z)$ are appropriately structured to produce all desired polarization states for the construction of 3D topologies. For the construction of a first-order skyrmionic hopfion beam, $E_R(r,\phi,z)$ and $E_L(r,\phi,z)$ must be in the form [78]

$$E_R(r,\phi,z) = (-A+B)LG_0^0(r,\phi,z) - BLG_1^0(r,\phi,z) \qquad (20)$$

and

$$E_L(r,\phi,z) = 2CLG_0^{-1}(r,\phi,z). \qquad (21)$$

The numerical values for parameters *A*, *B*, and *C* must be carefully selected in order to achieve uniform intensity in the beam cross-section and produce all desired polarization states in the hypersphere within the experimentally measured 3D volume of the skyrmionic hopfion beam. The azimuthal index-dependent gradient azimuthal phase of LG modes provides the uniform change in the azimuth of the polarization and the amplitude gradient (this being dependent on the radial index of the LG mode) in the radial direction produces the required gradient in the polarization ellipticity. Also, the order-dependent Gouy phase provides the desired propagation phase difference between the superposed LG modes. Hence, by controlling these three parameters in terms of radial and azimuthal indices of LG modes, we can obtain all phases and polarizations required to construct the optical hypersphere in a small volume of the light field. We can therefore generate localized optical hopfions in the light field.

First-order optical skyrmionic hopfions in the paraxial laser beam constructed by parameterizing its 3D topology structure in terms of polarization is depicted in Fig. 19. Fig. 19(a) depicts the polarization distribution in a finite volume of the laser beam (bounded by the beams transverse cross-section and propagation distance), obtained by the superposition of two orthogonally circular polarized light fields with suitable amplitude and phase distributions. Here, the uniform change in the transverse polarization distribution of the superposed laser modes depends on the amplitude and phase of the individual laser modes in the superposition. The transverse position of each polarization state on the beam cross-section smoothly changes as a function of propagation distance and without changing polarization state due to the Gouy phase difference created between the superposed laser modes under focusing conditions. Therefore, desired polarization loops for the fabrication of optical skyrmionic hopfions can be created in the confocal region. Each polarization fiber/filament is projected on a 2-sphere as a single point [Fig. 19(*b*)]. Optical fibration of polarization states along the beam propagation path can be clearly visualized in Fig. 19(*c*). In the present context, the polarization filament along the beam axis has RCP due to maximum intensity of $E_R(r,\phi,z)$ and zero intensity of $E_L(r,\phi,z)$ owing to a line singularity along its center. In the focal plane, we observe a dark ring in $E_R(r,\phi,z)$ with a finite intensity of $E_L(r,\phi,z)$ and a circular filament connecting LCP points. The Hopf fibration formed by all the polarization states becomes a nested torus structure in the volume containing the focal point of the paraxial laser beam Fig. 19(*c*).

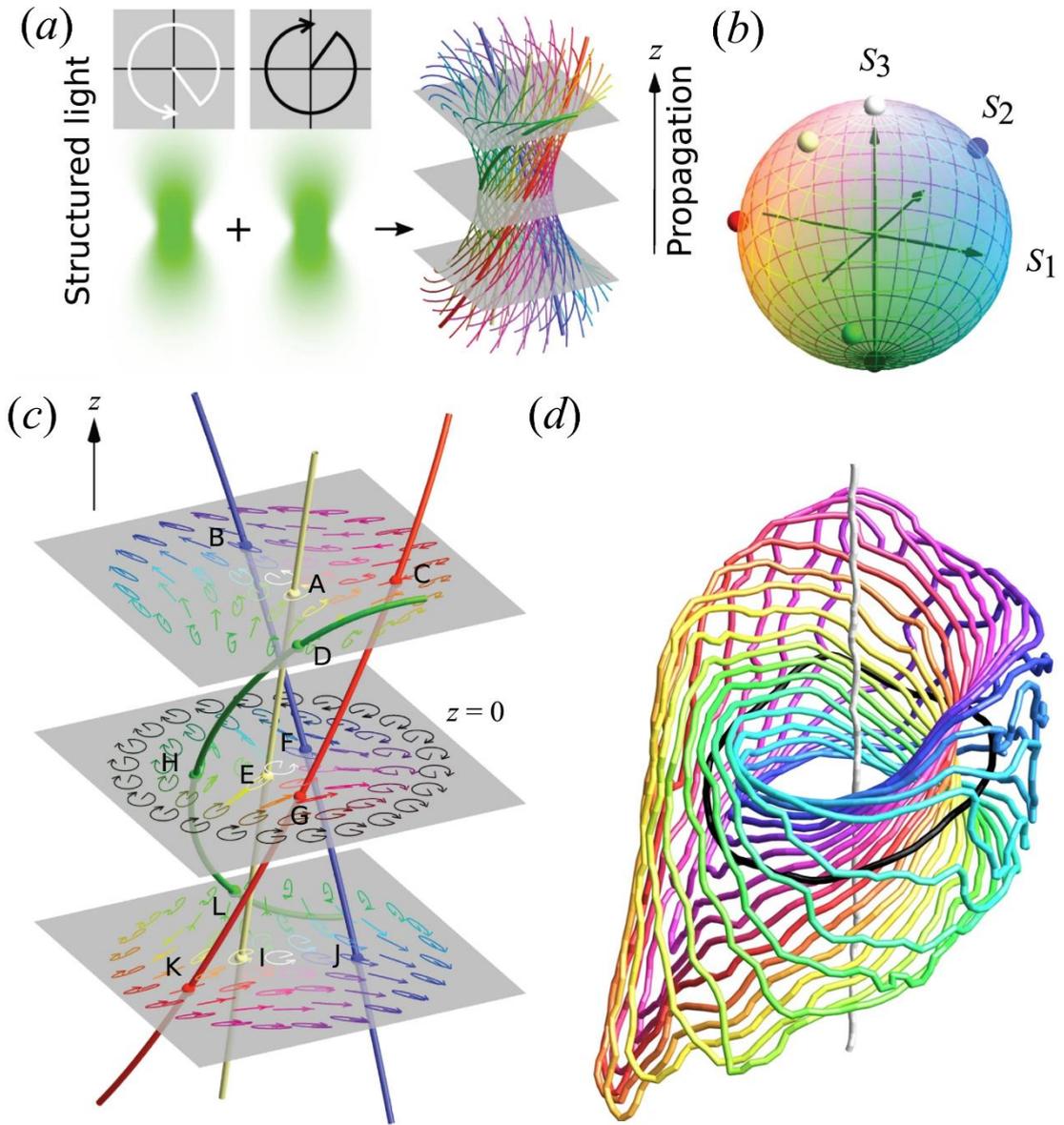

Fig. 19. Images detailing the characteristics of skyrmionic hopfions (3D quasiparticles) created in paraxial laser beams. (*a*) The superposition of two structured laser beams present in RCP (white) and LCP (black) produces polarization fibers which twist around the beam axis. Each fiber is formed by connecting the same elliptical polarization along the direction of propagation. To distinguish each fiber within the bundle, a gradient color was used with each fiber given a unique color based on polarization. (*b*) All the filaments can be represented on a single 2-sphere formed by Stokes vectors (Poincaré sphere). The north pole is RHP (white) and the south pole is (LHP). The polarization fibers correspond to four non-polar points on the 2-sphere and are plotted with their polarizations at three transverse planes in (*c*). (*d*) The optical texture in terms of Hopf fibration is constructed from the polarization and phase measurements of the skyrmionic hopfion beam. Here, the straight white line (RCP) and black circular loop (LCP) organize the texture into nested tori (obtained with permission from [78]).

In 2024, D. Ehrmanntraut, *et.al.* experimentally reported a second order skyrmionic hopfion by considering $E_R(r,\phi,z)$ and $E_L(r,\phi,z)$ as [80]

$$E_R(r,\phi,z) = (-D + 2E^4)LG_0^0(r,\phi,z) + 2E^4 LG_4^0(r,\phi,z) \tag{22}$$

and

$$E_L(r,\phi,z) = LG_0^2(r,\phi,z) \tag{23}$$

wherein parameters were optimized in the range provided by $D \geq 0$ and $E \geq 1$. The topology of the skyrmionic hopfion was constructed in paraxial laser beams distributed in the volume formed by the beam cross-section and propagation distance. The skyrmion density $\Sigma$ confined within the volume is given by [80]

$$\Sigma = \frac{1}{16\pi^2} \nabla\gamma \cdot (\nabla\cos\beta \times \nabla\alpha) \qquad (24)$$

Here, $\nabla$ denotes differentiation with regard to space, $-\pi < \alpha = \arctan(S_1, S_2) \leq \pi$ represents the polarization azimuth, $0 \leq \beta = \arctan(S_3) \leq \pi$ is the ellipticity, and $-2\pi < \gamma = \chi_R + \chi_L \leq 2\pi$ is the sum of the phases of the two constituent fields. $\chi_i$ being the phase of the respective right- or left-handed circularly polarized beam. To obtain the skyrmion number, a volumetric full-field reconstruction must be performed.

## 6. Experimental realization of optical quasiparticles

In the preceding sections, we theoretically examined the generation of optical quasiparticles in laser beams under the paraxial approximation through spatial modulation of polarization distributions. Polarization modulation was carried out in a controlled manner by superposing suitable generalized Gaussian modes. These optical quasiparticle textures can be experimentally generated in LG and Bessel modes and a flowchart showing the pathways to their generation is shown in Fig. 20. Two orthogonal generalized Gaussian modes present in two orthogonal polarization states can be created in a single laser beam either using the projection technique [81-83] or the interferometric technique [84-86].

By using two diffractive optical elements we can generate two orthogonal LG modes in laser beams and superpose them to generate optical quasiparticles in the LG profile. We can also generate Bessel quasiparticles by creating laser beams with an LG profile and HoG profile. By superposing these modes and passing them through a Bessel beam converter, we can generate optical quasiparticles in a Bessel profile. The state of these quasiparticles can be transferred between skyrmionic and bimeronic states by using a quarter-wave plate ($\lambda/4$) and a half-wave plate ($\lambda/2$). Skyrmionium and bimeronium quasiparticles can be generated by using nonzero radial index LG modes. Skyrmionic hopfions can be generated by the superposition of more than two orthogonal LG modes with suitable weighting factors.

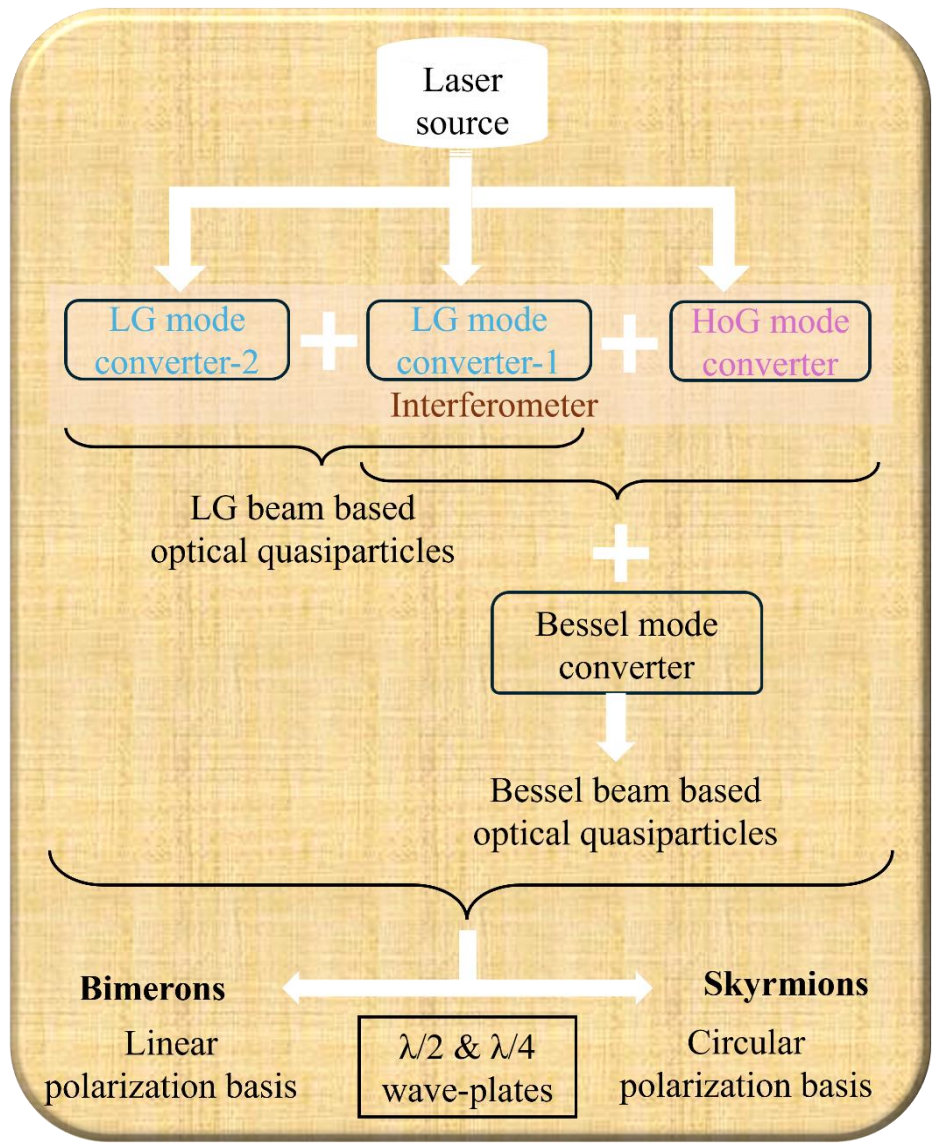

Fig. 20. Flowchart showing the pathways to quasiparticle generation in paraxial laser beams.

One of the simplest techniques to experimentally generate optical quasiparticles in paraxial laser beams makes use of a Mach-Zander interferometer (MZI). A MZI can be effectively used for most of the interferometric-based experimental configurations owing to its simplicity in alignment [87-89]. As shown in Fig. 21, a collimated Gaussian laser beam from a laser source can be split into two equal intensity laser beams which are orthogonally linear polarized by passing through a half-wave plate ($\lambda/2$) and the first Polarizing Beam Splitter ($PBS_1$). These two orthogonally oriented, linearly polarized modes can be superposed by $PBS_2$ to create a single laser beam which contains two orthogonal generalized Gaussian modes present with orthogonal polarization states. This can then be used to generate optical quasiparticles in paraxial laser beams.

***Optical quasiparticle generation in LG profile:*** By removing one of the mode converters from the MZI and replacing the other with an LG mode converting device (such as a SPP, SLM, or DMD), we can generate a superposition of Gaussian and LG modes at the output of the second PBS. The collinearly superposed modes will contain a tunable optical quasiparticle texture due to the presence of the $\lambda/2$ and $\lambda/4$ plates. It should be noted that the order of the quasiparticle is equal to the order of the LG mode generated in the

MZI. These quasiparticles are considered bright Poincaré beams owing to their bright central regions. We can also generate quasiparticles which are considered dark Poincaré beams by using two orthogonal LG mode converters in the place of mode converter-1 and mode converter-2. Critical here is that both mode converters must produce LG modes with unequal topological charges ($\ell_R \neq \ell_L$).

***Optical quasiparticle generation in Bessel profile:*** The same broad experimental MZI design as used with LG modes can be used to generate optical quasiparticles in Bessel profiles. In this case, mode converter-1 should take the form of a HoG mode converter, and mode converter-2, a LG mode converter. The state formed by superposing the HoG and LG modes is then passed through an axicon which is inserted next to the second PBS. This enables the generation of variable-order optical quasiparticles in the Bessel profile. It is also noted that we can generate optical quasiparticles with a central dark core by replacing the HoG mode converter with another LG mode converter. In the present experimental configuration, two orthogonal Bessel modes are generated from the single axicon and each have the same radial and longitudinal wave vectors. Thus, the two modes have the same Gouy phase and therefore will not facilitate transformations between Néel-type and Bloch-type quasiparticles. Instead, we can generate sequential transformations between Néel-type and Bloch-type quasiparticles by using two axicons with different apex angles instead of a single axion.

As shown in Fig. 21(*b*), multiple combinations can be used for the generation of propagation-dependent skyrmionic textures. In this experimental configuration, the input modes of the axion have an annular intensity distribution with a central dark core and this can be used to minimize deleterious effects of manufacturing defects which can often manifest at the tip (apex) of the axicon [90-92]. This can help avoid longitudinal intensity modulations in Bessel skyrmions. The mode conversion efficiency of this experimental configuration is optimal and minimizes power loss due to Fresnel reflections.

The experimental configuration shown in Fig. 21 can be further simplified by replacing the Gaussian laser and mode convertor combination with a vortex laser source [93-98]; in particular, vortex lasers wherein vortex modes are directly emitted from the laser cavity. Vortex modes are eigenmodes of the laser cavity and generally are of higher purity than vortex modes produced using external mode conversion techniques. Improved mode purity leads to the generation of higher quality skyrmions. Vortex lasers which directly generate vortex mode emission have seen a number of advancements in recent years. Notably sources with significant wavelength diversity (spanning the visible to near-infrared wavelength ranges) have been demonstrated [99-101]. By using such sources, it is anticipated that optical quasiparticles with broad wavelength diversity may also be realised.

The polarization basis of the superposed modes can be transferred to any arbitrary elliptical polarization basis by a pair of λ/2 and λ/4 plates and we can produce tunable bimeronic and skyrmionic textures in the laser beam. We can analyze the polarization distribution of skyrmions and bimerons by polarimetric analysis and projecting the polarization state on to the Stokes basis (Fig. 22 shows the Stokes vectors of first-order skyrmions and bimerons). The Stokes vector $S_0$ corresponds to the total intensity, and it is propagation-independent for skyrmions and bimerons. In skyrmions, $S_1$ and $S_2$ have petal structures and rotate with propagation. The number of petals created in the Stokes vector is double that of the skyrmion number. The vector $S_3$ is symmetric and invariant with propagation. In the case of bimerons, $S_1$ has circular symmetry due to the LP basis.

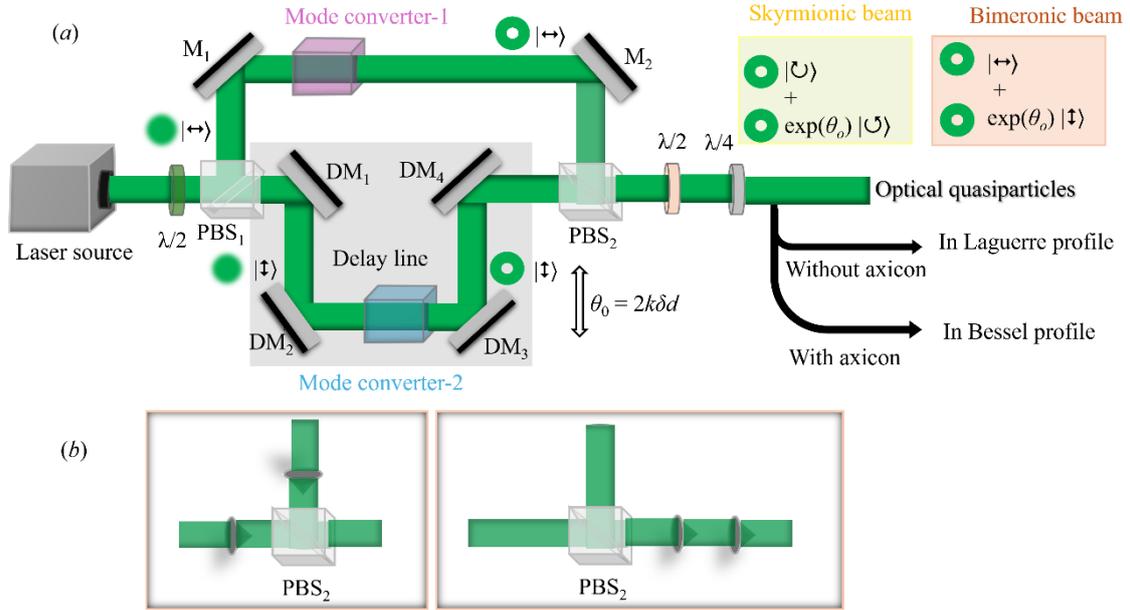

Fig. 21. (*a*) Schematic of the experimental, Mach-Zander interferometer based optical quasiparticle generation system. (*b*) The combination of axicons can be used for the generation of optical quasiparticles in Bessel profiles. $PBS_i$ is a polarizing beam splitter, $M_i$ is a mirror, $DM_i$ is the delay line mirror, $\lambda/2$ is a half-wave plate, and $\lambda/4$ is a quarter-wave plate. The horizontal and vertical linear polarization states are represented with respective $|\leftrightarrow\rangle$ and $|\updownarrow\rangle$ symbols. The right circular and left circular polarization states are indicated with $|\circlearrowright\rangle$ and $|\circlearrowleft\rangle$ symbols respectively.

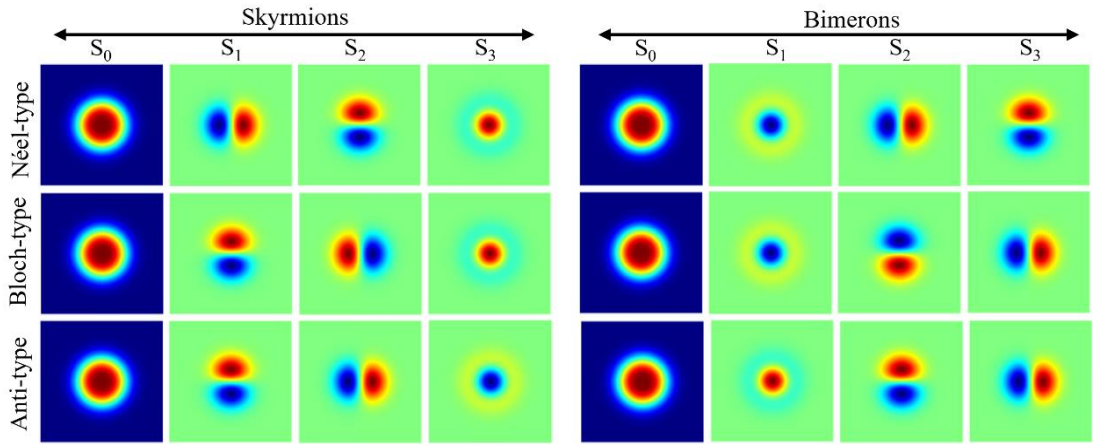

Fig. 22. Images showing the Stokes parameters of optical quasiparticles generated with LG modes.

As shown in Fig. 23, we can represent all possible states of skyrmions and bimerons generated from the experimental setup on a single torus. By using this experimental configuration which utilizes spiral phase plates and an axicon as mode converters, we can generate high-power LG and Bessel quasiparticles in both continuous wave (cw) and pulsed temporal regimes.

All possible textures of bright quasiparticles (table 2) can be generated using the above detailed experimental configuration [Fig. 24]. First, we can easily transfer the state of the quasiparticles smoothly between skyrmionic and anti-skyrmionic states either by simply interchanging the polarization states between the two orthogonal modes through wave plates or by changing the sign of the vortex mode in the presence of a dove prism. Furthermore, the transformation between Néel-type and Bloch-type quasiparticles can be achieved by providing a tunable phase delay $\theta_0 = 2k\delta d$ using the delay stage formed by the $DM_i$ mirrors in the MZI.

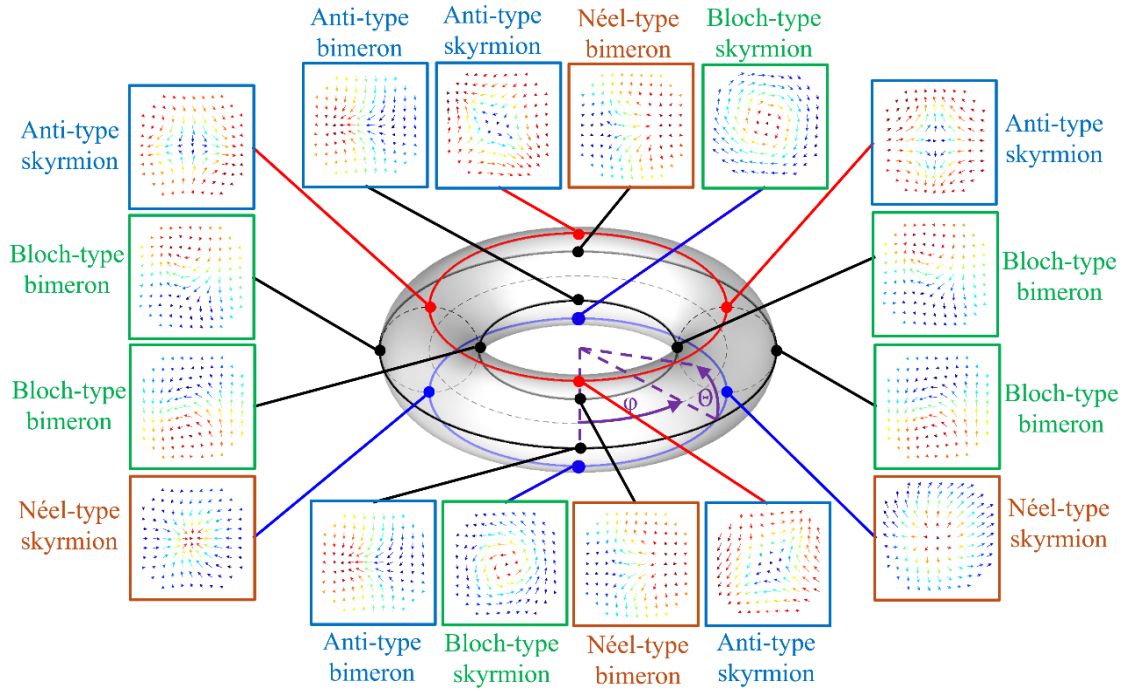

Fig. 23. First-order skyrmions and bimerons generated from a single experimental configuration as projected onto a torus.

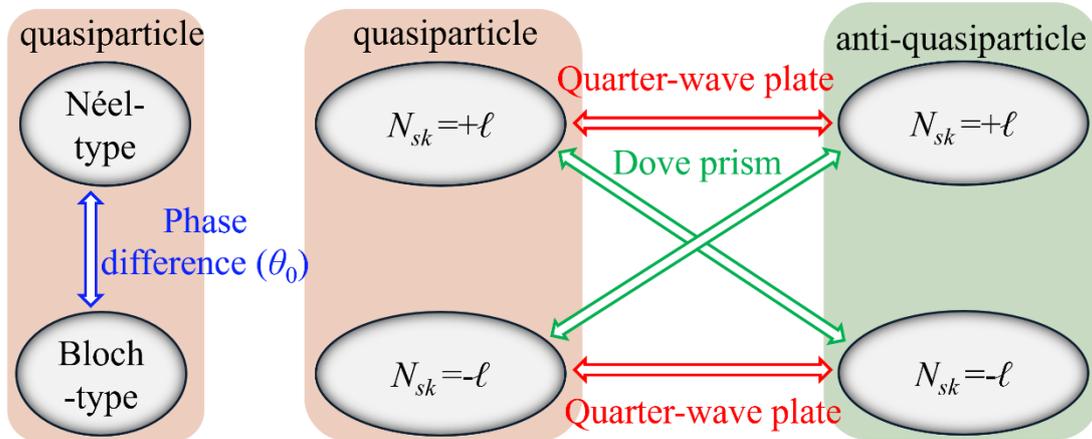

Fig. 24. Outline of how the transformation between various textures of arbitrary order bright quasiparticles generated in LG/Bessel profiles (obtained from [53]) takes place.

Recently, several techniques have been showcased for the generation of various kinds of optical quasiparticles in paraxial laser beams. Digital diffractive optical elements such as SLMs and DMDs have been used to generate the desired structured laser modes which were then interfered to produce tunable first order skyrmions in an LG profile [102]; order tunable skyrmions and bimerons [103]; first order [80], and second order [78] skyrmionic hopfions; Bessel-Gaussian optical skyrmions [52]; and meron and skyrmion lattices [65,104]. Recently, we utilized a non-interferometric technique based on a single SLM to generate order- and shape-tunable skyrmions [105].

Skyrmionic textures can also be generated in laser beams by using unconventional techniques such as Hall-effect-like splitting of vector beams to produce a pair of skyrmions [106]. Skyrmions have also been produced in pointing vectors by counter-propagating four vector beams under focusing conditions [18]. An array of optical Stokes skyrmions has been generated and tailored by tightly focusing cylindrical vector beams [107]. Accelerating particle-like polarization topologies and free-space bimeronic lattices have also been demonstrated by using an Airy beam [108].

In order to realize practical applications, quasiparticle laser sources need to be simple, compact and cost-effective. Additionally, it is highly desirable if they have the capacity to produce high order, shape tunable, and wavelength versatile outputs. Some of the techniques/approaches used to-date to generate these optical quasiparticles are summarized below:

**Spin-orbit wave plates:** Flat-optics skyrmion generators based on the spin-orbit interaction of light have been experimentally demonstrated. Here, spin-orbit wave plates directly transform a Gaussian beam into a skyrmionic beam [109].

**Photonic gradient-index lenses:** Vector structured light passing through gradient-index (GRIN) lenses has been shown to produce not only skyrmions but also complex topological textures like lattices of skyrmions and merons, so called skyrmioniums. Here, gradient birefringence is produced in a thin rod to support a Stokes skyrmionic mode as an eigenmode. An array of these thin rods are used to form a composite GRIN lens. In order to completely characterize the quasiparticles, the authors introduced multiple topological numbers (quantified based on centrality, radiality, vorticity, and polarity) in addition to the skyrmion number, to fully describe these photonic quasiparticles [110].

**Metafibers:** A metafiber is fabricated by sculpting a metasurface onto the tip of a fiber and this acts as an integrated photonic skyrmion generator. It can be tuned to output non-skyrmions and skyrmions by adjusting the polarization angle of the input/output light and is a method which confers a high degree of flexibility [111].

**Micro-ring cavity:** W. Lin, *et. al.* proposed and experimentally demonstrated skyrmion generation based on an optical micro-ring cavity [112,113]. As shown in Fig. 25(*a*), this was achieved by positioning two different angular gratings along with chiral lines on a micro-ring cavity. While the inner ring produced a RCP Gaussian beam, the outer ring produced a LCP vortex beam with the desired properties. Both of the modes generated from the concentric rings generated a full Poincaré beam because of the superposition of two orthogonal LG modes having the same Gaussian spot size and which were collinearly propagating. Thus, the output beam has a skyrmionic texture. The skyrmionic texture can be smoothly transformed between Néel-type and Bloch-type through rotation of one of the angular gratings with reference to the second one. Even though this laser system delivered very low output power, it is very compact and has the potential for use in micro-level applications.

**Conventional linear laser cavity:** Recently, we generated wavelength-tunable visible first-order quasiparticles (skyrmions and bimerons) from a linear laser cavity configuration [Fig. 25(*b*)]. A praseodymium solid-state laser (Pr: YLF) which has multiple emission lines in the visible spectrum was used [114-117]. The Gaussian and LG modes have different Eigen frequencies and as such cannot oscillate simultaneously in a conventional laser cavity. To overcome this, we first setup a linear laser cavity that produced a Gaussian mode and then we integrated a wedge shear interferometer (WSI) into the cavity. The WSI acted as a second output coupler and enabled the generation of a vortex mode without any effect on the laser cavities inherent operation [118,119]. The superposition of the generated Gaussian and vortex modes via a PBS resulted in the generation of skyrmions and bimerons under circular and linear polarization bases [120]. Due to the emission cross section of Pr: YLF, we were able to generate quasiparticles at wavelengths of 640 nm and 607 nm [121]. To increase the overall gain of the laser cavity we replaced the Pr: YLF laser crystal with a Pr: WPFG fiber and were able to produce multi-color quasiparticles (at wavelengths of 523, 605, 637, and 719 nm) [67]. In this process, the Gaussian mode was generated from the laser cavity and the vortex mode was generated from a WPSI and had a high mode purity of >97 %. The quasiparticles generated using this technique had high spatial and temporal stability and were topologically protected under propagation.

We anticipate that by further modifying the laser cavity, optical quasiparticles at other wavelengths may be produced. Here, other solid-state gain media may be used as well as non-linear wavelength conversion techniques [122-124]. The WSI used was made of glass and could operate over a broad wavelength range and also had a high damage threshold. Thus, it should facilitate broad spectral operation and high power handling. Overall, the laser cavity design is simple, compact, and cost-effective.

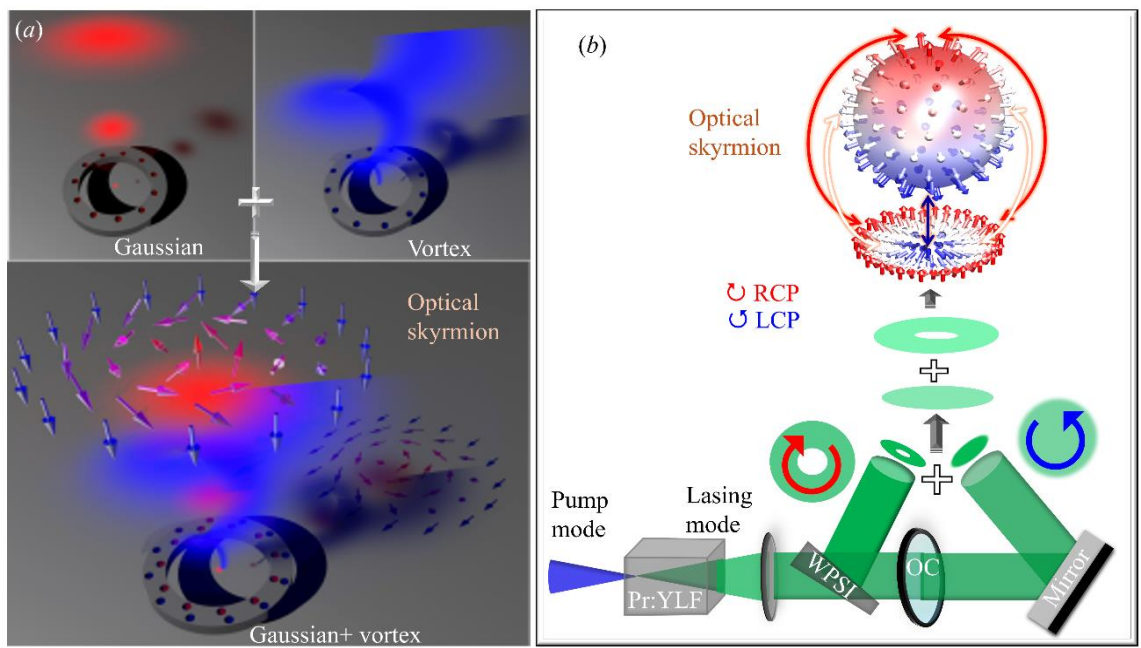

Fig. 25. Images outlining the principle behind compact quasiparticle laser sources. (*a*) Skyrmion generation using micro-cavities; and (*b*) dual output coupler-based praseodymium laser source generating skyrmions and bimerons in the visible wavelength range Fig. 25(*a*) is obtained with permission from [113].

## 7. Wavelength tuning of optical quasiparticles

Over the last few decades there has been significant research into the development of laser systems capable of generating emission across a plethora of wavelengths spanning the ultra-violet to mid-infrared [125-127]. Typically, the lasing wavelength of a laser systems is dependent on the characteristics of the lasing medium; however, the incorporation of nonlinear optical processes can be used to tune this wavelength for specific applications [128]. Nonlinear wavelength conversion processes are typically phase matched and as such, require specific orientations of the propagation vector and polarization of the light field. The wavelength of linearly polarized laser beams can be up/down-converted by nonlinear conversion processes and can be applied to scalar beams [129,130].

Vector beams, in contrast, have an inhomogeneous polarization distribution across their cross section and it is therefore not possible to simultaneously phase match all of their polarization states. However, this can be somewhat overcome through independent frequency doubling of the orthogonally LP components in the MZI [Fig. 21(*a*)] formed using two PBS. For example, by inserting two orthogonally-oriented, identical second harmonic generation (SHG) crystals in the two arms (in the place of mode converters) of a MZI for frequency doubling. By inserting suitable half-wave and quarter-wave plates, we can generate the desired vector beam at the lasers second harmonic wavelength. The SHG system can be further simplified (and made more robust) by replacing the MZI with a Sagnac interferometer and by using a single SHG crystal [131,132].

In the case of much shorter and longer wavelengths, PBSs and mirrors may not be available or be exceedingly expensive, and the above technique may not be suitable for the generation of vector beams at new wavelengths. An alternative technique based on two orthogonally oriented nonlinear crystals under a single pass configuration [133] can be used to generate vector beams at new wavelengths in terms of LP basis. We can generate these vector modes in a circular polarization basis only if there is availability of quarter-wave plates at said new wavelengths. By using the above techniques under the superposition of orthogonal LG/Bessel modes, we can produce optical quasiparticles at the new desired wavelengths (Fig. 26). As mentioned in the previous sections, we can transform the texture of quasiparticles between skyrmionic and bimeronic states by using wave plates. For frequency conversion, quasiparticles must be in a bimeronic state where we have two orthogonal LP states. By inserting two orthogonally oriented nonlinear crystals about their phase matching axis in the bimeronic beam, we can frequency double its wavelength. It is worth noting that the SHG process doubles the order of fundamental optical quasiparticles. The second harmonic bimeron can be transformed into a skyrmion by using a quarter-wave plate. A major drawback of wavelength tuning through the use of nonlinear optical processes is that the quality of the quasiparticles degrades and the process is typically of low conversion efficiency. We can increase the conversion efficiency by focusing the quasiparticle beams onto nonlinear crystals, however, the effects of spatial walk-off and saturation increase with the increase in focusing, and the quality of the generated modes further degrade [134-137].

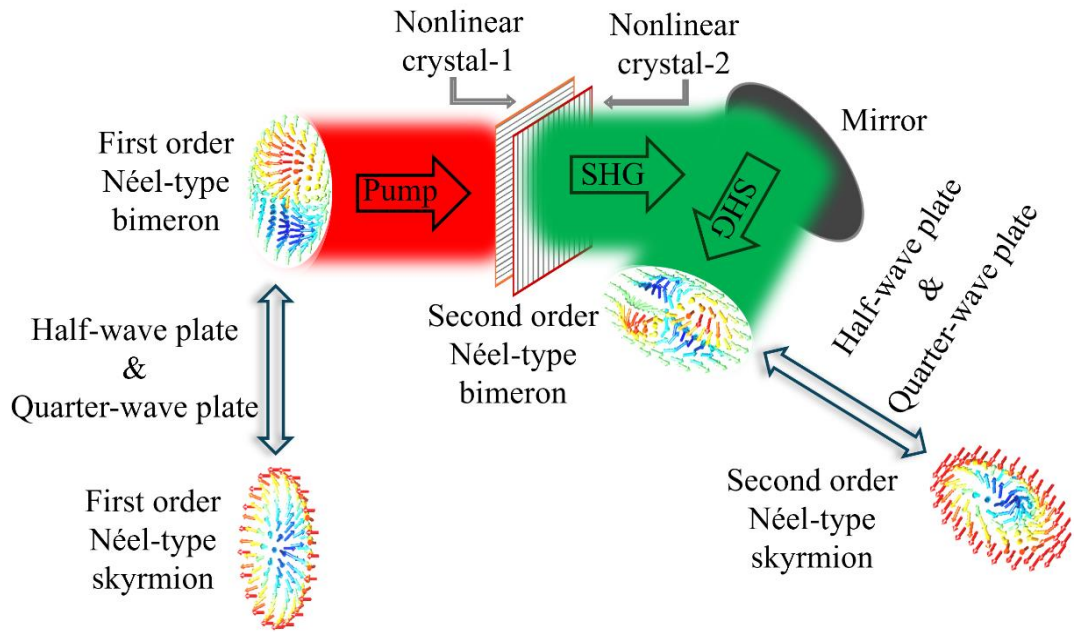

Fig. 26. Image showing the principle by which second harmonic generation (frequency-doubling) of optical quasiparticles can be achieved in paraxial laser beams.

## 8. Experimental artefacts in the optical quasiparticles of paraxial laser beams

Depending on the pump source, cavity configuration and output power, some laser systems produce low-quality Gaussian modes and when these pass through a mode converter, will result in the generation of a structured mode with low mode quality. Similarly, low quality modes may also be generated through the combination of high-quality Gaussian outputs from a laser system and the use of diffractive optical elements of low quality (due to artifacts and limitations of their manufacture). Ultimately, the quality of generated optical quasiparticles depends on the mode purity of the superposed structured modes. Therefore, the generation of high-quality, propagation-independent topological textures in paraxial laser beams required the use of a high-quality Gaussian beam and excellent-quality mode converters. Even though we have demonstrated several techniques for the experimental generation of quasiparticles (discussed in the experiment section), depending on the desired wavelength, we must select an appropriate technique that can produce optimum-quality quasiparticles. Optical quasiparticle generation and characterization in paraxial laser beams is a hot-topic and has only gained momentum in recent years. We anticipate that a plethora of new experimental techniques for the generation of these quasiparticles will emerge imminently.

The topological texture of quasiparticles in paraxial laser beams is experimentally characterized by analyzing the polarization distribution in the beams cross-section. In the case of an LG mode with a large outer radius and a polarization distribution as shown in Fig. 27(*a*), we can easily estimate the skyrmion number (close to flat intensity distribution). However, this is not true in real-world experiments. Here, the polarization is characterized in terms of intensity, which decreases with increasing radial distance (in LG and Bessel modes). Hence, the polarization distribution of real/experimental quasiparticles take the form shown in Fig. 27(*b*) and Fig. 27(*c*). The CP components present in the peripheral region have a very low intensity and the signal-to-noise ratio decreases with increasing radial distance. As a result, it affects the skyrmion number calculation [Fig. 27(*d*)]. The area of selection for the theoretical calculation of skyrmion numbers can be easily fixed and we can obtain integer skyrmion numbers. In the case of experimental data, background noise and the intensity distribution near the periphery of the mode restricts the area over which/from which the skyrmion number can be determined [103]. Therefore, careful selection of the area over which/from which the skyrmion number is determined from experimental data, is crucial.

One method by which the background noise (mostly in the form of stray light) can be reduced is by characterizing the polarization distribution under dark lab conditions. It is also important to note that the experimental skyrmion number depends on what kind of polarimetric technique is used and also the quality of the wave-plates used in the measurement.

The quality of the quasiparticles in the paraxial laser beams depends on the intensity of superposed generalized Gaussian beams in their transverse spatial overlap region. For example, in LG mode quasiparticles, the effective overlap between a vortex mode with a Gaussian mode depends on the order of the vortex mode [Fig. 27(*e*)]. The overlap area effectively decreasing with increasing vortex mode order. Theoretically, it has no effect on the skyrmion number calculation because all the Stokes vectors are normalized with their local intensities and there are no external sources. However, it is not true in case of experimental realization. Extracting the polarization states at low intensity overlapping is difficult for higher-order optical quasiparticles. As this overlap decreases, it becomes progressively more difficult to produce all of the polarization states of the Poincaré sphere in the beam cross-section; which is especially important when generating higher-order optical quasiparticles. This hence places a limit on the experimental generation of real-world higher-order optical quasiparticles in paraxial laser beams.

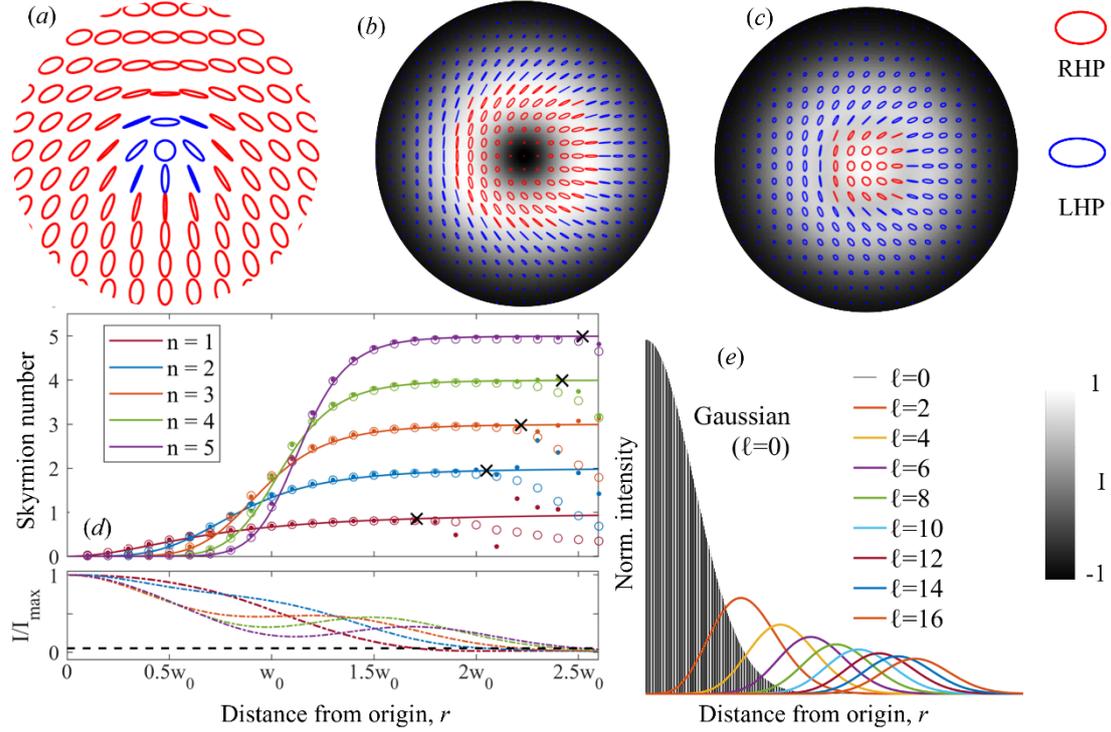

Fig. 27. Images showing (*a*) the polarization distribution a skyrmion. The polarization distribution of a dark skyrmion is shown in (b) and that of a bright skyrmion in (c), both superposed onto the spatial intensity distribution of the beam (the size of the elliptical polarization depends on the intensity). (*d*) Plot of the skyrmion number as a function of radial position (solid lines are numerically modelled data and filled and open circles are values estimated from experimental data). (*e*) Plot of the transverse spatial overlap between a Gaussian mode with variable order vortex modes as would be used for skyrmion generation. The Gaussian mode area is shaded in black. Fig. 27(*d*) is replicated with permission from [103].

Another important characteristic is the stability of a quasiparticle as it undergoes propagation. As shown in Fig. 28, when we focus superposed modes at different propagation distances, a propagation-dependent mode mismatch will manifest which leads to the degradation of the topological texture [5]. This scenario can occur in experimental systems when the focusing optics have spherical aberration [38,138-140]. Spherical aberration focuses the Gaussian beam farther than LG modes and can degrade the quality of the superposed modes [141-143]. In the practical application of quasiparticles, spherical aberration of the focusing system must hence be minimized. Undesired mixed modes can occur in the superposed laser beams due to laser sources, nonlinear wave-mixing, and diffractive optical elements and produce propagation-dependent impurities in the topological texture of optical quasiparticles. Hence, an understanding of the propagation stability of optical quasiparticles is required prior to utilizing them in applications.

The stability of quasiparticle textures depends on their polarization distribution. Depolarization effects due to diffractive optical elements and propagation media can degrade the quality of topological textures. Also, quasiparticles which reflect from any surface, transform into anti-quasiparticles and vice versa.

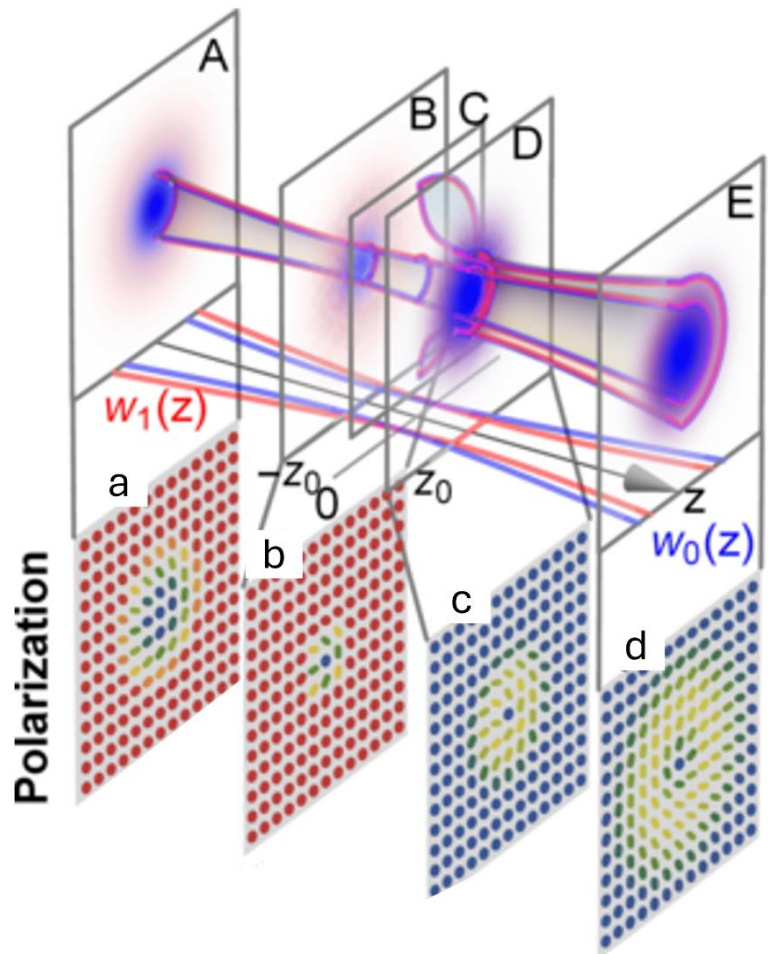

Fig. 28. Image showing the propagation characteristics of a first-order skyrmion formed by the superposition of LG modes focused at different distances. Panels A-E show the intensity distribution, and panels a-d show the corresponding polarization distribution of the beam across its cross-section and at different propagation distances (obtained with permission from [5]).

## 9. Current and future prospects for optical quasiparticles

The major advantage of paraxial quasiparticles with reference to other quasiparticles is that they can be easily synthesized and characterized. These modes can propagate without distortion in free space and linear media and provide more high degrees of freedom which can be exploited across a plethora of applications.

To-date, optical quasiparticles in paraxial laser beams have been demonstrated in the visible and near-infrared wavelength ranges (as covered in the experiment section) and we anticipate that this may be extended to other regions of the electromagnetic spectrum as experimental techniques and methodologies evolve. To the best of our knowledge, there have been no experimental reports of the generation of single meron textures in Stokes vectors. We expect that such experimental demonstrations will come to the fore as the superposition of different structured beams is undertaken.

As discussed in the preceding sections, substantial work has been carried out on paraxial optical quasiparticle generation and characterization, however, there have been relatively few reports on their applications. In the near future, we expect to see several applications of these quasiparticles across multiple research fields. The non-diffracting and self-healing topological textures created through Bessel quasiparticles can have unique polarization-based applications in light-matter interactions and can provide further insights into fundamental and applied sciences.

## 9.1. Light-matter interaction

Topologically structured scalar and vector laser beams can be used to fabricate unique structures on the micro- and nano-scale on materials. Also, structured light can be used as a probing tool for materials characterization. The topologically protected polarization textures of 2D and 3D skyrmions can be used for the fabrication of a diversity of structures with subwavelength dimensions in polarization-sensitive materials [144-146]. Recently, we fabricated the 3D texture of skyrmionic hopfions onto micron-thick azopolymer films with the beam undergoing different propagation distances [147]. Here, the polarization textures of optical hopfions produced surface relief structures via optical radiation pressure (Fig. 29).

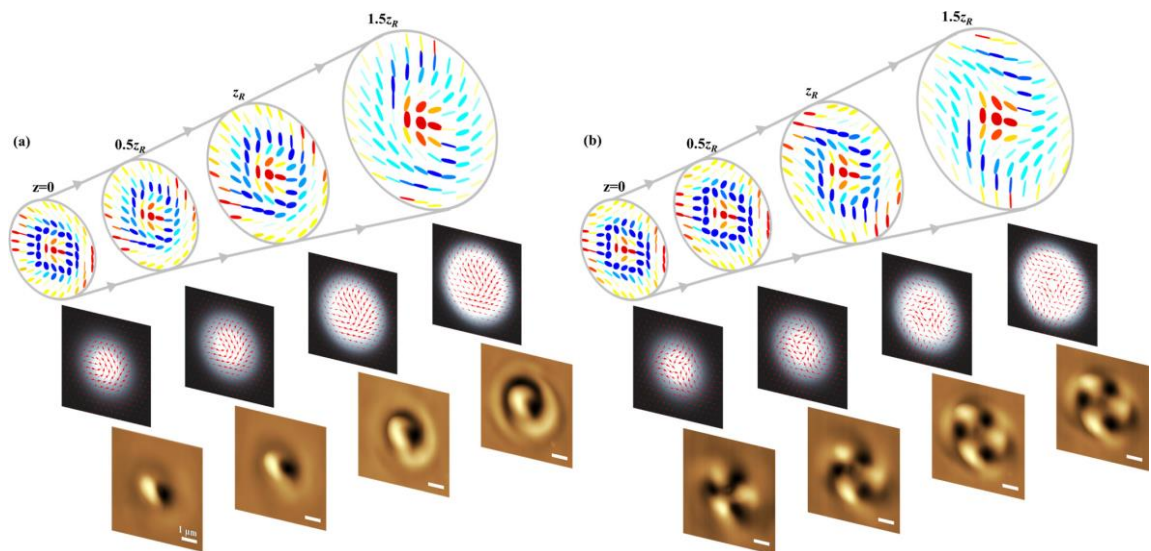

Fig. 29. Images showing the 3-dimensional projection of the polarization textures (theoretically modelled), time-averaged optical scattering forces, and experimentally obtained surface relief structures (in azopolymer film) by the illumination of (*a*) Néel-type (*b*) Bloch-type skyrmionic hopfions (reproduced from [147]).

Further advancements in the field of optical quasiparticles include the demonstration of polarization Möbius strips, 3D polarization distributions, and spin-orbit interactions under tight focusing of 2D paraxial quasiparticles; such demonstrations may enable new applications in material science and advanced engineering technologies [148-152]. In the case of such beams and under tight focusing conditions, the z-component of the electric field is utilized in applications including trapping, imaging, and materials processing. The 3D polarization can be experimentally traced in the cylindrical volume created by the transverse plane (*xy*-plane) and the propagation axis (*z*-axis). The resultant paths created by the polarization have a 3D topology and the created shapes are called optical knots, or optical threads. We can fabricate these knots in different shapes. For instance, torus knots and links, the figure-eight knots, etc.

In the future, it is critical that we investigate the polarization components (s-polarization, p-polarization, transverse polarization, and longitudinal polarization) [153-158] of these quasiparticles under tight focusing conditions in order to better our understanding of new structures which can be created in the confocal region and maximize their use in novel light-matter interactions.

## 9.2. Optical communications

Due to propagation-dependent polarization distributions, some Poincaré beams are not ideal for use in optical communications. However, this case is not the case for Poincaré beams which have propagation-independent, stable quasiparticle textures. In the case of paraxial quasiparticles, even though their textures created in paraxial laser beams transform from one state to another state, the skyrmion number is invariant in linear, homogeneous, and isotropic media, a characteristic which is ideal for use in optical communications. The polarization texture is also invariant, even when they propagate through complex media [159,160]. In order to understand the topological textures' stability while the quasiparticle propagates through linear and nonlinear mediums, in [161], a theoretical model was developed which considers the evolution of the Stokes parameters, polarization singularities, and optical skyrmionic textures in a medium. They showed that the skyrmion tube formed by a central C-line surrounded by an L-surface has a longitudinally invariant topological number. Recently, a generalized skyrmion number was introduced in [162]. In their study, when the skyrmionic beam propagated through various types of media, the skyrmion number was not preserved due to the existence of singularities and the lack of appropriate symmetries of the boundary. However, the newly defined generalized skyrmion number shows excellent robustness and remains constant with propagation. Also, as discussed in the experiment section, we can easily synthesize optical quasiparticles with a desired skyrmion number. With respect to application in optical communications, the desired wavelength of quasiparticles [163-165] can be obtained through the use of nonlinear-wave-mixing techniques (enabling coverage of wavelength bands 780-850 nm & 1520-1600 nm). For underwater optical communications (where the ideal wavelength band is 450-550 nm), praseodymium laser-based optical quasiparticles can be used.

Quantum entanglement based optical communication is a secure communications technique which has been demonstrated with the use of optical paraxial skyrmions [166,167]. Quantum entanglement can be produced in these quasiparticles in different ways: entanglement of local skyrmions, entanglement of non-local skyrmions, and non-local skyrmion entanglement of local skyrmionic photons. The topology of skyrmionic beams provides non-local quantum entangled states which are robust to smooth deformations of the wave function and remain intact until the entanglement itself vanishes. It should be noted that the background noise and signal-to-noise ratio have no correction while the entangled states have correlation and the decryption of encrypted data through quantum entanglement is free from noise. Further, the loss in the transmission path and noise in the detection process can be reduced by properly selecting the detection basis [168]. It has also recently been proposed that perturbation-resilient integer arithmetic can be performed with skyrmion numbers through the use of passive optical components [169].

### 9.3. Imaging

The structural properties of living and non-living organisms can be visualized by using optical microscopes with resolution limited by the wavelength of the illuminating light source and the therapeutic window. This limitation has been overcome by using nonlinear optical process and structured light beam illumination [58,170-173]. Nonlinear optical imaging techniques have superior signal-to-noise ratio and have been able to suppress background noise arising from the pump light. However, the spatial resolution is still limited to $1/\sqrt{2}$ times the diffraction limit, as determined by the wavelength of the illuminating light. In structured light imaging, first-order optical vortex-based super-resolution imaging was successfully achieved by stimulated emission depletion (STED) [174,175] and up-conversion fluorescence depletion (UFD) techniques [176-178]. In addition to the above techniques, the topologically protected inhomogeneous polarization texture of skyrmions may enable polarization-based high spatial resolution imaging via nonlinear light-matter interactions.

### 10. Summary and conclusions

A quasiparticle texture can only exist in paraxial laser beams if the polarization and the field amplitude are spatially varying; Poincaré beams meet these requirements. Optical quasiparticles in paraxial laser beams are derived from Poincaré beams but not all Poincaré beams are quasiparticles. Optical quasiparticles are quantitatively characterized by their skyrmion number (also referred to as the order of the quasiparticle). The skyrmion number must be an integer and its sign can be negative or positive. Experimentally obtained skyrmion numbers may deviate from this (not be an integer) due to experimental artifacts and constraints. The skyrmion number describes the number of times the tip of the Stokes vector covers the entire Poincaré sphere. Quasiparticles are classified into three types Néel-type, Bloch-type, and anti-type.

Paraxial vector laser beams have topological texture in their Stokes vector field, and by examining their topological textures, we can confirm whether the quasiparticle texture belongs to Néel-type or Bloch-type or anti-type. It is worth to note that the azimuthal-dependent polarization decides what kind of skyrmionic texture can have the paraxial Poincaré beam.

Within the field of photonics, the topological textures of quasiparticles with tunable shape and order can be relatively easily fabricated and characterized in paraxial laser beams, with a high degree of cost-effectiveness. Quasiparticles fabricated in most research areas are confined in the 2D plane and only propagate in the transverse direction with reference to their texture. However, in the case of quasiparticles in paraxial laser beams, they propagate in the direction perpendicular to their textures in free space and other media. Therefore, optical skyrmions are free-space topological textures and we can readily utilize them in many applications. The propagation direction of optical quasiparticles can be easily controlled, and their topology can be transformed using low cost processes/techniques. Optical skyrmions can be generated using a range of parameters and those which are not based on a Stokes vector field, are either confined in very small regions or form in a particular plane and change their topology with propagation [178-183]. Optical skyrmions generated in paraxial laser beams in general, have propagation-independent topological textures.

The quasiparticle textures discussed in this review are generated within Laguerre-Gaussian (LG) and Bessel beams under the paraxial approximation. Skyrmions and bimerons with an arbitrary order can be easily generated in both LG and Bessel beams. The amplitude of the skyrmion number (order of the quasiparticle) depends on the difference in the OAM numbers of the superposed structured laser beams. The sign of the skyrmion number depends on the permutation and combination between superposed structured modes and vectors in a polarization basis. The superposition of structured modes in an elliptical polarization basis produces bimerons; and these bimerons become skyrmions for the special case of a circular polarization basis. LG quasiparticles are diffractive and preserve their topological texture throughout their propagation. Bessel quasiparticles have a non-diffracting and a self-healing nature. It is also noted that skyrmionic hopfions are 3D topological particles while the rest of the optical quasiparticles discussed in this review are 2D quasiparticles. We experimentally demonstrate skyrmionic hopfions which are generated and characterized within a LG basis.

Optical quasiparticles in paraxial laser beams are experimentally generated via the superposition of orthogonally-structured modes which have two orthogonal polarization states. The skyrmionic textures of these quasiparticles can be easily characterized by classical Stokes polarimetry. The transformation between Néel-type and Bloch-type skyrmions can be achieved by simply changing the phase difference between the superposed LG modes. A transformation between skyrmionic and anti-skyrmionic textures can be made by interchanging the polarization states of superposed structured modes or by changing the sign of their OAM number. When LG quasiparticles are passed through focusing optics, they transform their topological texture between Néel-type and Bloch-type without changing their topology.

In the case of Bessel skyrmions, we can generate propagation-dependent periodic transformations between Néel-type and Bloch-type quasiparticles by superposition of Bessel beams with different radial propagation vectors. Also, when superposed Bessel beams have the same radial propagation vectors, no transformation between Néel-type and Bloch-type topological textures will take place. In the case of LG quasiparticle generation, we look at only transverse mode matching, however, in the case of Bessel quasiparticles, we must mode match both transverse and longitudinal modes for the generation of non-diffracting and self-healing topological textures with stable propagation characteristics.

In addition to a single optical quasiparticle, we can also generate a lattice of quasiparticles in a single laser beam. While a single quasiparticle has a circular boundary, quasiparticle lattices have a non-circular boundary. The polarization geometry of these lattices depends on the characteristics of the interfered, generalized Gaussian modes. The individual quasiparticles share their boundary with neighbouring quasiparticles and have a skyrmion number less than their integer value. The stereographic projection of lattice quasiparticles on a 2-sphere is completely different than conventional, single quasiparticle projection.

By incorporating nonlinear optical processes, we open up the possibility of generating optical quasiparticles across a range of different wavelengths, and which may be tuned for specific applications. To-date, optical skyrmions have been utilized across applications including quantum entanglement and topologically protected polarization printing on azopolymers. Optical quasiparticles in paraxial laser beams are robust and resilient to atmospheric turbulence and exhibit very stable propagation characteristics in complex media. As such, they hold significant promise as an enabling technology which will open new vistas in fields which extend beyond that of structured material fabrication and optical communications.